\documentclass[acmsmall]{acmart}

\AtBeginDocument{%
  \providecommand\BibTeX{{%
    \normalfont B\kern-0.5em{\scshape i\kern-0.25em b}\kern-0.8em\TeX}}}

\setcopyright{acmcopyright}
\copyrightyear{2018}
\acmYear{2018}
\acmDOI{XXXXXXX.XXXXXXX}

\acmJournal{JACM}
\acmVolume{37}
\acmNumber{4}
\acmArticle{111}
\acmMonth{8}

\graphicspath{{figures/}}
\usepackage{graphicx}
\usepackage{multirow}
\usepackage{booktabs}
\usepackage{threeparttable}
\usepackage{pifont}
\usepackage{wasysym}
\usepackage{makecell}
\usepackage{bm}

\makeatletter
\usepackage{comment}
\let\wfs@comment@comment\comment
\let\comment\@undefined
\usepackage{changes}
\let\wfs@changes@comment\comment
\let\comment\@undefined
\newcommand\comment{%
    \ifthenelse{\equal{\@currenvir}{comment}}
    {\wfs@comment@comment}
    {\wfs@changes@comment}%
}
\makeatother

\begin{document}

\title{A Survey on Machine Unlearning: Techniques and New Emerged Privacy Risks}


\author{Hengzhu Liu}
\email{hzliugrimm@stu.zuel.edu.cn}

\author{Ping Xiong}
\authornotemark[1]
\email{pingxiong@zuel.edu.cn}
\affiliation{%
  \institution{Zhongnan University of Economics and Law}
  \streetaddress{182\# Nanhu Avenue, East Lake High-tech Development Zone}
  \city{Wuhan}
  \country{China}
  \postcode{430073}
}

\author{Tianqing Zhu}
\affiliation{%
  \institution{University of Technology Sydney}
  \streetaddress{123 Broadway}
  \country{Australia}
  \postcode{Ultimo NSW 2007}
}
\email{Tianqing.Zhu@uts.edu.au}

\author{Philip S. Yu}
\affiliation{%
 \institution{University of Illinois at Chicago}
 \streetaddress{1200 W Harrison St}
 \city{Chicago}
 \country{United States}
 \postcode{Illinois 60607}
}
\email{psyu@uic.edu}

\renewcommand{\shortauthors}{H. Liu et al.}

\begin{abstract}
The explosive growth of machine learning has made it a critical infrastructure in the era of artificial intelligence. 
The extensive use of data poses a significant threat to individual privacy.
Various countries have implemented corresponding laws, such as GDPR,
to protect individuals' data privacy and the right to be forgotten.
This has made machine unlearning a research hotspot in the field of privacy protection in recent years,
with the aim of efficiently removing the contribution and impact of individual data from trained models.
The research in academia on machine unlearning has continuously enriched its theoretical foundation,
and many methods have been proposed, targeting different data removal requests in various application scenarios.
However, recently researchers have found potential privacy leakages of various of machine unlearning approaches, making the privacy preservation on machine unlearning area a critical topic. 
This paper provides an overview and analysis of the existing research on machine unlearning,
aiming to present the current vulnerabilities of machine unlearning approaches. 
We analyze privacy risks in various aspects, including definitions, implementation methods, and real-world applications.
Compared to existing reviews,
we analyze the new challenges posed by the latest malicious attack techniques on machine unlearning from the perspective of privacy threats.
We hope that this survey can provide an initial but comprehensive discussion on this new emerging area.
	
\end{abstract}


\maketitle

\section{Introduction}\label{intro}

Due to the important role of machine learning algorithms, 
abundant computer systems are beginning to hold a large amount of personal data for decision-making and management.
For example, the increasingly notable ChatGPT actively utilizes big amount of datasets for knowledge discovery~\cite{Wang2023survey}.
However,
research has shown that machine learning models can remember information about training data \cite{Feldman2020,Carlini2019},
raising concerns about potential attacks on individual privacy.
Adversarial attacks,
such as membership inference attacks \cite{Salem2018} and model inversion \cite{Fredrikson2015},
have demonstrated the ability to extract information about target data from machine learning models.
In response to these concerns,
there have been significant developments in regulations and laws governing individual privacy.
For instance,
the General Data Protection Regulation (GDPR) implemented by the European Union \cite{Mantelero2013},
and the California Consumer Privacy Act (CCPA) specifically state \emph{the right to be forgotten}.
The data owners are then obligated to respond to these deletion requests promptly \cite{Chen2021gan} and lead to a new technology emerging: machine unlearning~\cite{Cao2015}. 

Machine Unlearning
refers to the process of removing both the data and its influence on a machine learning model.
A straightforward way is to achieve machine unlearning is by retraining the model from scratch.
However,
this can result in significant computational time and overhead,
particularly when dealing with large datasets and models with complex structures.
Consequently,
the crucial challenge in machine learning is to address the issue of how to mitigate the impact of data that must be forgotten on the model without a complete retraining process.

Numerous studies have been conducted on the subject of machine unlearning,
and existing techniques fall mainly into one of two types: data-oriented unlearning and model-oriented unlearning~\cite{Xu2024}.
Data-oriented methods implement data removal by modifying the original training set.
For example,
Bourtoule et al. \cite{Bourtoule2021} introduced an unlearning method called \emph{SISA},
which involves dividing the training set into several disjoint subsets and training each submodel independently.
By retraining the submodel, the
data can be effectively erased.
Model-oriented methods involve manipulating the original model,
like parameters,
to achieve unlearning.
Such as the Certified Removal mechanism,
proposed by Guo et al. \cite{Guo2020},
aims to offset the influence of forgotten data on model and update the model parameters.

\begin{figure}[h]
	\centering
	\includegraphics[scale=1]{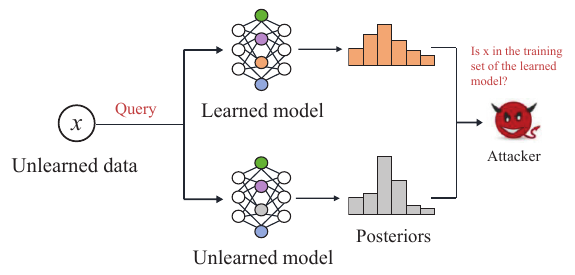}
	\caption{The Example of Privacy Violation in Machine Unlearning}
	\label{over}
\end{figure}

Even though the target of machine unlearning is to protect the privacy of the data,
recent researchers have found that unlearning schemes may jeopardize privacy in an unexpected way.
This is because the majority of current machine unlearning schemes have applied the machine learning algorithms, therefore inherited the natural weaknesses or flaws of those learning algorithms.  
Chen et al.~\cite{Chen2021when} have pointed out that the difference between the output distributions of two models before and after unlearning can lead to additional information leakage of forgotten data, which greatly challenges the design of unlearning algorithms.
For example, in data-driven schemes, 
partitioning the original training set enables to improve efficiency by retraining the affected submodels.
In this way,
the prediction results of retrained submodels are reflected in the confidence vectors through the aggregation operation,
resulting in the differences between outputs of the learned and unlearned models,
which may cause privacy violation that exposes membership information about the unlearned data.
When an attacker queries the learned and unlearned models,
the differences between posteriors of the unlearned data can be obtained by the attacker,
which are used to determine whether the unlearned data is a member of the learned model's training set.
This process is shown in Fig. \ref{over}.

There is no clear definition for the privacy risk of machine unlearning.
Current surveys have not fully presented the privacy issues of current machine unlearning schemes.
Utilizing machine unlearning needs further analysis and summary, especially when it comes to privacy attack and protection.
Therefore,
we believe that it is timely for a more comprehensive and systematic survey of privacy issues of machine unlearning.
Through summarizing and comparing current unlearning schemes,
we can have a deeper understanding of advantages and disadvantages of each branch of methods,
and performance evaluation and real-world application of unlearning are also involved to identify future research directions.
These series of analysis can help current researchers, model owners, and data owners to be aware of the existing privacy risks of machine unlearning scheme, while alert the potential privacy risks in the future.


In addition to highlight the privacy threats and discuss the vulnerability of each unlearning scheme,
we provide a clear taxonomy of unlearning methods with potential privacy risks, and summarize the potential defenses,
along with the risks of prevalent applications.
The contributions of this survey are listed as follows:

\begin{itemize}
	\item{We proposed a taxonomy that classify current machine unlearning methods into two branches, data-oriented and model-oriented techniques.}
	\item{We carefully analyzed the privacy risks in unlearning based on attack schemes, and highlighted the vulnerability of different unlearning techniques. To tackle those privacy violations, we presented possible defense methods.}
	\item{We reviewed the applications of machine unlearning and potential privacy issues. In addition, we discussed on the current research trends and principal issues, while identifying potential future directions.}
\end{itemize}


\section{Preliminaries}\label{preliminaries}

\subsection{Notations}

In a machine learning task based on a supervised learning setting,
$\mathcal X$ and $\mathcal Y$ represent the instance space and label space, respectively.
The training set $D$ can be defined as $D=\{(\bm{{\rm x_1}}, y_1), (\bm{{\rm x_2}}, y_2), \ldots, (\bm{{\rm x_n}}, \\ y_n)\}$,
where $\bm{{\rm x_i}} \in \mathcal X$,
and $y_i \in \mathcal Y$ which is the label of $\bm{{\rm x_i}}$. A machine learning model, $M_o$ with parameters $w_o$, is originally trained from the training set $D$
by a machine learning algorithm $A$,
namely,
$M_o = A(D)$.
In machine unlearning,
let $D_f$ be the data samples to be forgotten,
and $D_r=D\setminus D_f$ be the retained set with the forgotten sample deleted.
Taking the original model $M_o$, the data set $D$ and $D_f$ as inputs,
an unlearning algorithm $U$ aims to generate a new model, $M_u$, that has the similar performance with $M_r$,
where $M_r$ is the model trained on $D_r$.
$w_u$ and $w_r$ represents the parameters of the unlearned model $M_u$ and retrained model $M_r$,
respectively.
We denote the distribution of the models $M_r= A(D_r)$ by $\mathcal P_{r}$,
and the distribution of the models $M_u=U(M_o,D,D_f)$ by $\mathcal P_{u}$.
The two distributions are ideally expected to be identical. We list the above notations and their corresponding descriptions in
Table~\ref{table_1}.


\begin{table}[!h]
	\begin{center}
		\caption{Notations}\label{table_1}
		\begin{tabular}{c|l}
        \hline
	\toprule
			Notation & Description \\
	\midrule
			$D$ & Original training dataset\\
			$D^{'}$ & New training dataset\\
			$D_f$ & The data that needs to be forgotten\\
			$D_{r}$ & Retain training dataset\\
			$D_{S}$ & The subset of training dataset\\
			$M_o$ & The original model    \\
			$M_r$ & The retrained model    \\
			$M_u$ & The unlearned model    \\
			$\mathcal X$ & The instance space \\
			$\mathcal Y$ & The label space \\
			$\bm{{\rm x_i}}$ & A sample in $D$ \\
			$y_i$ & The label of sample $\bm{{\rm x_i}}$ \\
            $(\bm{{\rm x_n}}, y_n)$ & A sample with features in training set \\
			$A$ & Machine learning algorithm\\
			$U$ & Machine unlearning algorithm\\
			$\mathcal P_{r}$ & Distribution of the retrained model\\
			$\mathcal P_{u}$ & Distribution of the unlearned model\\
			$w_o$ & The parameters of original model \\
   		  $w_r$ & The parameters of retrained model \\
			$w_u$ & The parameters of unlearned model \\
	\bottomrule
		\end{tabular}
	\end{center}
\end{table}

\subsection{Background of Machine Unlearning}

While machine learning discovers general knowledge
from historical data,
machine unlearning removes identified samples and their contributions from a learning model.
Considering that large-scale machine learning tasks are extremely expensive in terms of
time complexity and computational cost,
fully retraining the model from the scratch is impractical.
Thus,
we expect to achieve machine unlearning and
obtain the unlearned model $M_u$ efficiently without fully retraining on dataset $D_r$.
In addition,
the unlearned model $M_u$ should have a similar performance with that of the retrained model $M_r$.
In another word,
the distribution of unlearned models and retrained models
should have the identical or indistinguishable distributions.
The formal definition of machine unlearning is as follows:

\begin{definition}[Machine Unlearning \cite{Bourtoule2021}]\label{MUL}
	Let ${M_{o}}$ be the original model trained on $D$,
	$D_f$ be the data samples to be forgotten.
	A machine unlearning algorithm takes ${M_{o}}$ and $D_f$ as inputs and outputs the unlearned model ${M_{u}}$.
	Suppose ${M_{r}}$ be the retrained model learned from $D_r=D \setminus D_f$,
	we say $D_f$ is unlearned from ${M_{o}}$ if $\mathcal P_{r}$ and $\mathcal P_{u}$ are identical or indistinguishable, namely
	\begin{equation}\label{lma}
	\mathcal P_{r} \cong _P \mathcal P_{u}
	\end{equation}
	where $\mathcal P_{r}$ and $\mathcal P_{u}$ are the distributions of $M_{u}$ and $M_{r}$.
\end{definition}

\begin{figure}[h]
	\centering
	\includegraphics[scale=0.5]{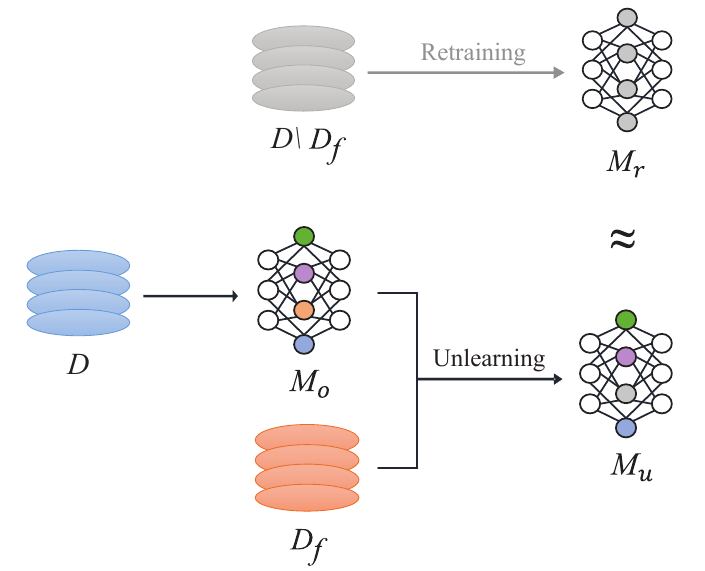}
	\caption{The Objective of Machine Unlearning}
	\label{mulframe}
\end{figure}

Fig. \ref{mulframe} illustrates the objective of machine unlearning.
Give the forgotten data $D_f$,
we can obtain the retrained model $M_r$ learned from $D_r=D \setminus D_f$.
The unlearned scheme produces model $M_u$ by using 
$M_o$ trained on $D$ and $D_f$.
The purpose of machine unlearning scheme is to make $M_u$ indistinguishable from $M_r$.

\subsection{Techniques of Machine Unlearning}

\begin{figure}[h]
	\centering
	\includegraphics[scale=1]{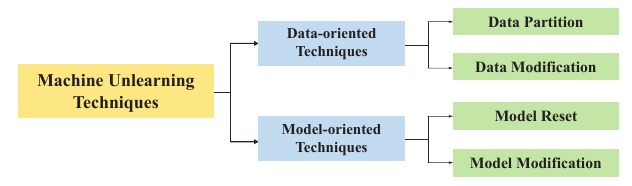}
	\caption{The Techniques of Machine Unlearning}
	\label{unlearningtype}
\end{figure}

Based on the revising targets, we categorize the machine unlearning approaches into data-oriented and model-oriented techniques.
The former is unlearning methods that manipulate the training set,
including data partition and data modification,
while the latter is unlearning methods that modify the original model,
including model reset and model modification. 
Fig. \ref{unlearningtype} summarizes the taxonomy of machine unlearning techniques.

\subsubsection{Data-oriented Techniques}

Data-oriented techniques refer to the unlearning methods that data holders erase the forgotten data through manipulating the original training set $D$,
mainly including two types according to different processing strategies,
namely data partition and data modification.

\textbf{Data Partition:}
In data partition,
the data holders divide the original training set into several subsets and train the corresponding submodels on each subset.
Then these submodels are used to aggregate a prediction using an aggregation function.
When an unlearning request is received,
the data holders delete the forgotten data from the subsets that contain them and retrain the correspoding submodels.

\textbf{Data Modification:}
In data modification,
the data holders modify the training set $D$ with adding noise or new transformed data $D_{T}$,
e.g. using transformed data $D_{T}$ to replace the forgotten data $D_f$,
that is $D^{'} = D_r \cup D_{T}$.
For example,
Cao et al. \cite{Cao2015} converted the training set into a summation form,
which is easy to compute,
so that the unlearning process only depend on updating the part of summation which affected by $D_f$.
This type of methods can simplify retraining and speed up unlearning.

\subsubsection{Model-oriented Techniques}

Model-oriented techniques refer to the unlearning methods that data holders complete unlearning through modifying the original model ${M_{o}}$.
The technique can be categorized into model reset and model modification.

\textbf{Model Reset:}
In model reset,
the data holders directly update the model parameters to eliminate the impact of forgotten data $D_f$ on the model ${M_{o}}$,
that is, $w_u = w_o + \sigma$,
where $w_o$ and $\sigma$ are the parameters of ${M_{o}}$ and the value to update,
respectively.
Certified removal proposed by Guo et al. \cite{Guo2020} is the fundamental form of model reset.
Its basic idea is that calculating the influence of $D_f$ on model and updating the corresponding parameters.
But these approaches usually increase computational costs,
especially for models with complex structures.

\textbf{Model Modification:}
In model modification,
the data holders replaces the relevant parameters of $D_f$ with the calculated parameters to remove $D_f$ from the original model ${M_{o}}$,
that is,
$w_u = w_o \cup w_{cal}$,
where $w_{cal}$ is the calculated parameters for replacement.
Taking the unlearning method proposed by Brophy et al. \cite{Brophy2021} as an example.
This type of methods are usually suitable for machine learning models and deep learning models with simple structures, 
like random forests, 
as they require pre-calculated parameters.

\subsection{Privacy Risk of Machine Learning and Unlearning}


\subsubsection{Taxonomy of Privacy risks in Machine Learning}

\begin{figure}[h]
	\centering
	\includegraphics[scale=1]{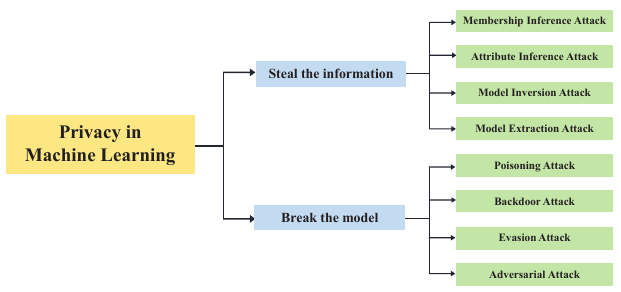}
	\caption{The Taxonomy of Privacy in Machine Learning}
	\label{privacyml}
\end{figure}

We first analyze the widely recognized privacy attacks on machine learning from the perspective of attack targets \cite{Liu2018survey, Liu2021when}. 
The machine learning can be categorized into the following two types, as illustrated in Fig. \ref{privacyml}.

\textbf{To steal the information:}
Users want to ensure the privacy of training data when using a machine learning model,
including data values,
features,
or whether a certain data belongs to the training set (membership).
Additionally, 
due to the ability to replicate trained models, 
model structure and parameters are also considered as sensitive information.
This private information on training data and models is the main target of attacks,
which typically includes membership inference attacks, attribute inference attacks, model inversion attacks, and model extraction attacks.

\textbf{To break the model:}
The privacy attacks aimed at breaking the model primarily refer to the destroy of the model's integrity,
which is the accuracy of model predictions compared to the expected outputs \cite{Xu2019}. This type of attacks on machine learning mainly includes poisoning attacks,
evasion attacks,
adversarial attacks and backdoor attacks,
resulting in a negative impact on the model performance,
such as reducing accuracy, effectiveness, and efficiency. 

\subsubsection{Privacy in Machine Unlearning}

The original purpose of machine unlearning is to prevent privacy disclosures that may arise as a result of machine learning,
thereby safeguarding personal data privacy.
However, 
previous studies have indicated that machine unlearning also presents potential privacy risks \cite{Ellers2019, Chen2021when}.
Since the removal of forgotten data and its impact on the model, 
machine unlearning techniques modify the original machine learning model, 
which may lead to additional privacy disclosure issues.

There are two main reasons that contribute to privacy concerns in machine unlearning.
The first one arises from the discrepancy between the learned model and unlearned model.
The process of unlearning results in generating two different models,
which can exhibit differences in prediction outputs,
parameters,
gradient updates and other related aspects,
which might contain the sensitive information of the forgotten data.
The attacker can utilize inconsistencies in model outputs to conduct attacks to acquire sensitive data.
This potential privacy threat is referred to as information-stealing attacks in this paper.
Specifically,
information-stealing attacks include membership inference attack,
attribute inference attack, 
model inversion attack and model extraction.
Existing research \cite{Chen2021when} has demonstrated that the two outputs of the model generated by the \emph{SISA} unlearning framework can leak membership information about forgotten data through membership inference attacks.


Another one is the vulnerability caused by data deletion operation itself,
which can be used to compromise the model's integrity and availability.
The former aims to introduce a backdoor to compromise the unlearned model's prediction capability through backdoor attacks.
Due to the necessity of data deletion,
the attacker can carefully modify the training data,
such as data labels,
then initiates a series of malicious unlearning requests to trigger the hidden backdoor,
misleading the unlearned model into making incorrect predictions over specific samples to destroy the utility of model \cite{Hu2023, Zhang2023backdoor}.
The latter occurs within the framework of approximate unlearning,
which relies on updating gradients or intermediate parameters, 
resulting in the approximation error.
When this error exceeds the predefined threshold, 
complete retraining will be activated to satisfy the indistinguishability.
This limitation provides attackers with an opportunity to poison the training data to increase computational overhead \cite{Marchant2022}.
Such issues may cause that the information the model unlearned exceeds what it ought to unlearn,
slowing down the unlearning process and degrading the unlearned model's performance.
reduces the prediction accuracy of the unlearned model.
We classify these phenomenons as model-breaking attacks.

Based on the above reasons for the potential privacy issues in machine unlearning,
we categorize the privacy threats of unlearning into two types,
information-stealing attacks and model-breaking attacks.
Membership inference attack,
attribute inference attack,
model inversion attack and model extraction attack are fall into information-stealing attacks,
while model-breaking attacks mainly include backdoor attack and poisoning attack.

\begin{figure}[h]
	\centering
	\includegraphics[scale=1]{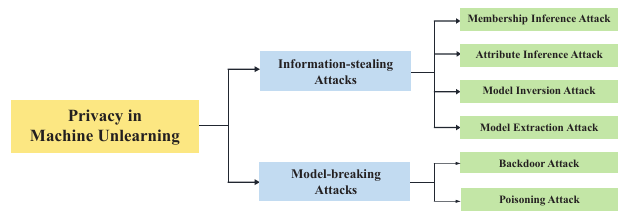}
	\caption{The Taxonomy of Privacy in Machine Unlearning}
	\label{privacymul}
\end{figure}

Fig. \ref{privacymul} shows the taxonomy of privacy issues of machine unlearning. We divided the attacks based on the targets. Please note that in the model-breaking attacks, even though the poisoning attacks are considered as security attack in some literature, as the potential risks can lead to information leakage, we still include it as privacy risk in this survey. The two types of privacy threats of machine unlearning are discussed in detail in Section \ref{privacythreat}.

\subsubsection{Comparison of Privacy in Machine Learning and Unlearning}

Machine learning and machine unlearning both involve privacy risks,
with similarities and differences between them.
On the one hand,
machine learning and unlearning face issues of data privacy and model privacy.
Data privacy includes data value (e.g. recovering an image),
membership and certain attributes,
while model privacy includes model parameters,
training algorithms and availability.
Therefore,
the privacy attacks for learning and unlearning are similar,
which can be used to steal information and break models.

However,
on the other hand,
machine learning and unlearning are differed in terms of attack goals, targets and workflows.
Firstly,
although both privacy attacks on learning and unlearning aim to extract information and destroy model,
the specific attack goals of each method are different.
For example,
membership inference attack, attribute inference attack, and model inversion attack on machine learning are designed to extract information of the training data, 
while in machine unlearning, 
they aim to steal the unlearned data's privacy information.
Model extraction attacks are used to replicate the trained models, 
but for unlearning, 
it can clone the behavior of the learned model to obtain the output of unlearned data.
Backdoor attack and poisoning attack break the unlearned model's utility.

Secondly,
privacy attacks in machine learning mainly aim at the training data and trained models.
However,
in unlearning,
attack targets include the unlearned data,
learned models and unlearned models.

Finally,
from the perspective of attack workflows,
the outputs of a machine learning model,
such as the predicted label and class probability vector,
can be used to construct shadow models to steal private information,
including data membership and model parameters.
The outputs of the two versions of models,
the learned model and the unlearned model,
generated by machine unlearning,
could be combined to create feature vectors, as the differences between them may expose information about the unlearned data.
A comparison of privacy in machine learning and unlearning is shown in Table \ref{comparisonmlmul}.

\begin{table*}[!h] \scriptsize
	\centering
	\caption{Comparison of Privacy in Learning and Unlearning}
	\label{comparisonmlmul}
	\begin{threeparttable}
		\resizebox{\textwidth}{!}{
		\begin{tabular}{cc|ccc|ccc}
			\hline
   \toprule
			\multicolumn{2}{c|}{\multirow{2}{*}{Privacy attacks}} & \multicolumn{3}{c|}{Machine learning} & \multicolumn{3}{c}{Machine unlearning} \\ \cline{3-8}
			\multicolumn{2}{c|}{} & Goals & Targets & Workflow & Goals & Targets & Workflow \\ 
   \midrule
			\multirow{4}{*}{\makecell[c]{Information\\-stealing}} & \makecell[l]{Membership \\inference \\attack} & \makecell[l]{Extracts membership \\ information of \\training data} & \makecell[l]{Training\\ data} & \multirow{4}{*}{\makecell[l]{Constructs\\ attack model\\ using outputs \\ of the \\trained model}} & \makecell[l]{Extracts membership \\ information of \\unlearned data} & \makecell[l]{Unlearned\\ data} & \multirow{4}{*}{\makecell[l]{Constructs\\ attack model\\ using outputs \\ of the \\learned model \\and unlearned \\model}} \\ \cline{2-4} \cline{6-7}
			& \makecell[l]{Attribute \\inference \\attack} & \makecell[l]{Extracts attribute \\information of \\training data} & \makecell[l]{Training\\ data} &  & \makecell[l]{Extracts attribute \\ information of \\unlearned data} & \makecell[l]{Unlearned\\ data} &  \\ \cline{2-4} \cline{6-7}
			& \makecell[l]{Model \\inversion \\attack} & \makecell[l]{Reconstructs the \\training data} & \makecell[l]{Training\\ data} &  & \makecell[l]{Reconstructs the \\unlearned data} & \makecell[l]{Unlearned\\ data} &  \\ \cline{2-4} \cline{6-7}
			& \makecell[l]{Model \\extraction \\attack} & \makecell[l]{Copies the \\trained model} & \makecell[l]{Trained \\ model} & & \makecell[l]{Copies the \\learned model} & \makecell[l]{Learned \\ model} &  \\ \hline
			\multirow{4}{*}{\makecell[c]{Model\\-breaking}} & \makecell[l]{Backdoor \\attack} & \makecell[l]{Destroys the\\integrity of \\model} & \makecell[l]{Trained \\ model} & \makecell[l]{Injects backdoor \\triggers} & \makecell[l]{Destroys the\\integrity of \\model} & \makecell[l]{Unlearned \\ model} & \makecell[l]{Injects backdoor \\triggers and\\requests unlearning} \\ \cline{2-5} \cline{6-8}
            & \makecell[l]{Poisoning \\attack} & \makecell[l]{Destroys the \\availability of\\model} & \makecell[l]{Trained \\ model} & \makecell[l]{Poisons the\\ training data} & \makecell[l]{Destroys the \\availability of\\model} & \makecell[l]{Unlearned \\ model} & \makecell[l]{Poisons the \\training data} \\ 
   \bottomrule
		\end{tabular}
	}
	\end{threeparttable}
\end{table*}

\section{Machine Unlearning Techniques and Vulnerabilities}\label{multech}


\subsection{Data-oriented Techniques}


\subsubsection{Data Partition}

Data partition involves dividing the original training set into several subsets and training submodels on each subset.
These submodels are then used to aggregate the prediction results.
Data partitioning unlearning methods can be categorized into two types based on the different data structures:
unlearning for linear data and unlearning for nonlinear data.


\paragraph{Unlearning for Euclidean Space Data}

Euclidean space data is a typical data structure, 
such as text and images.
The widely used exact unlearning approach for Euclidean space data is to partition the training set and conduct partial retraining.
Bourtoule et al. \cite{Bourtoule2021} proposed such a general method,
named \emph{SISA} (sharded, isolated, sliced, and aggregated learning).
The core idea of \emph{SISA} is to divide the training dataset into several disjoint subsets and train sub-models on each subset,
respectively.
When a specific data point needs to be unlearned, 
only the corresponding submodel requires retraining.

\begin{figure}[h]
	\centering
	\includegraphics[scale=0.75]{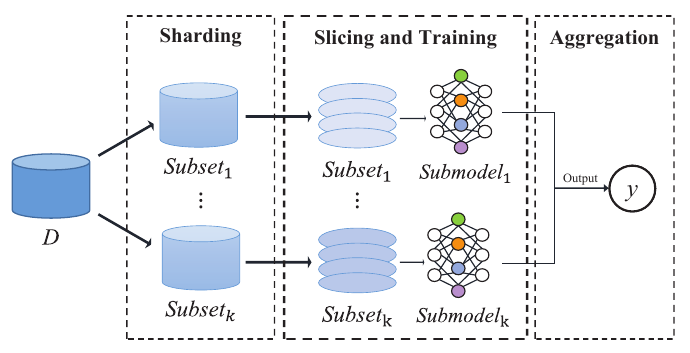}
	\caption{SISA}
	\label{SISA}
\end{figure}

Fig. \ref{SISA} shows that SISA consists of three stages.
i) \emph{Sharding}.
In this stage,
the training set $D$ is uniformly partitioned into $k$ disjoint subsets,
Each subset $D_S$ is thus used for training a corresponding submodel.
ii) \emph{Slicing and training}
Each subset $D_{S}$ is further uniformly partitioned into $r$ disjoint slices,
Then training process of the submodel $\mathcal{M}_S$ is performed by a number of rounds,
while only one slice is involved as the training set in the beginning round,
and one more slice is added to the training set
in each of the subsequent rounds.
The state of parameters associated with each round is saved for potential usage in machine unlearning.
Therefore,
given a data-point $d_u$ to be unlearned,
the retraining process only needs to begin from the round that the slice containing $d_u$ is added to the training set.
iii) \emph{Aggregation}.
It refers to using the predictions from all the submodels to aggregate an overall prediction.
A specific strategy is that each submodel provides an equal contribution by label-based majority vote.

In data partition technologies, there are other researchers work on how to partition and grouping data.
Focusing on exact data deletion requests,
Ginart et al. \cite{Ginart2019}
defined data deletion as an online problem~\cite{Bottou1998} and further proposed
a deletion efficient learning algorithms for $k$-means clustering,
Divide-and-Conquer $k$-means (DC-k-means).
This method uses a tree hierarchy to restrict the influence of model parameters to specific data partitions by dividing the training set.
Then in the process of data deletion,
this algorithm only needs to recompute the centroid of the specific partition which contained the data point to be deleted,
and recursively merge the results to achieve unlearning.
The DC-$k$-means unlearning algorithms provide a deletion efficient solution for large-scale clustering problems,
which improve the efficiency of machine unlearning.
Nevertheless,
these algorithms simplify the assumption of application scenarios,
such as considering the condition of only a single model,
and limiting data deletion requests to one data point.

In practical application scenarios,
unlearning may involve adaptive removal requests based on the published models.
In model inversion attacks~\cite{Veale2018},
specific deletion requests are made to update the published models,
Additionally,
if the training dataset has been compromised by adversarial attacks~\cite{chakraborty2018adversarial} where malicious samples have been intentionally implanted, 
these samples must be removed sequentially to gradually improve the published model.
Gupta et al. \cite{Gupta2021} defined the above unlearning requests as adaptive removal requests.
They proposed an adaptive machine unlearning strategy based on a variant of SISA
along with differential privacy and max-information \cite{Brown2021},
which reduces adaptive deletion guarantees to those of non-adaptive deletion \cite{dwork2015generalization}.
They also provided a strong deletion guarantees for non-adaptive deletion sequences to ensure the effectiveness of the unlearning process.

\paragraph{Unlearning for Non-Euclidean Space Data}

Unlike common Euclidean space data, 
non-Euclidean space data usually exhibit certain relationships,
such as graphs.
A graph consists of a set of vertices (V) and a set of edges (E), denoted by G(V, E).
Thus, 
unlearning requests in the context of graph include node unlearning and edge unlearning.
To deal with these unlearning requests,
Chen et al.~\cite{Chen2022} proposed a machine unlearning framework for graph data, GraphEraser,
following the basic pipeline of SISA.
GraphEraser consists of three phases, graph partition, shard model training
and shard model aggregation.
Specifically,
a graph is firstly partitioned into a set of disjoint shards (sub-graphs).
A shard model is then trained with each of the shard graphs.
During the aggregation phase,
the predictions of all the shard models are aggregated as the final result,
for example, by majority voting.
Consequently,
upon unlearning a node or an edge,
only the shard model corresponding to the unlearned node/edge
needs to be retrained.


Wang et al. \cite{Wang2023inductive} proposed a guided inductive graph unlearning to solve the problems of data removal in model agnostic inductive graph.
This framework consists of three components.
The first one involves guided graph partition with fairness and balance,
achieved by optimizing a spectral clustering problem.
Next,
it repairs subgraph after completing the partition by generating missing neighbors for each node based on its features and original degree information.
Finally,
aggregating all subgraphs according to the importance score,
which is computed by similarity between associated subgraph and the test graph.
Although it provides advantageous performance and efficient unlearning,
it increases computation cost due to the repair operations.

\subsubsection{Data Modification}

Data modification is a type of unlearning techniques that data holders can manipulate the original training set with transforming data or adding noise.
The data modification approaches mainly include two types based on the different modification strategies:
unlearning via data transforming and noise adding.

\paragraph{Unlearning via Data Transforming}

This type of methods transform the training data into a computable form,
and only updates a part of transformation.
The statistical query unlearning algorithm proposed by Cao et al. \cite{Cao2015} is a typical example of this type of data modification techniques.
The core idea is transforming the training set $D$ into a summation form,
instead of directly training on $D$.
These summation forms are computable transformations,
to make the training process independent on the training set,
but relies on summations.
When unlearning forgotten data $D_f$,
only the summations affected by $D_f$ need to be updated,
and then recalculating the model.
This unlearning method can conveniently unlearn data, 
but,
it is only applicable to models capable of conducting statistical query learning,
like Bayesian models,
SVM and so on,
meaning that it is not well-suited for large-scale models with complex structure~\cite{Rawat2022}.


Similarly, 
researchers have explored using transformed data to achieve attribute-level unlearning.
Guo et al. \cite{Guo2022} proposed attribute unlearning,
which contributes to improve model performance.
They formalized the problem of attribute unlearning as the cooperation between a truthful service provider and data holder.
This unlearning framework can be divided into two parts,
a representation detachment extractor and a classifier.
The detachment extractor is used to extract intermediate feature representations from the training process,
and uses their mutual information to measure the correlation between attributes,
so that sensitive attributes can be separated.
Then,
attributes that need to be forgotten can be removed by transforming them into representation forms and iteratively updating the model.
This attribute-level unlearning scheme can preserve model fidelity with high efficiency.

\paragraph{Unlearning via Noise Adding}

Another type of data modification is to perturb the training data by adding noise,
which destroys the information about unlearned data to achieve data erasure.
Tarun et al. \cite{Tarun2021} proposed an unlearning framework with error-maximizing noise generation and impair-repair based weight manipulation,
which allows to unlearn a class of data samples from a machine learning model.
This method first generates an error-maximizing noise matrix for the unlearned class,
and uses the noise matrix with a high learning rate to 'impair' the weight of the class or classes to achieve class-unlearning.
Moreover,
since such a manipulation may disturb the weights of retained classes,
they introduced the 'repair' step to regain the performance through training the machine learning model for a single epoch.
Overall,
this method provides excellent class-unlearning and retains the accuracy of model.
It can also be extend to solve problems with large dataset as few update steps are required for unlearning multiple classes.

\subsubsection{Discussion and Potential Vulnerabilities of Data-oriented Techniques}

It is worth pointing out that nearly all data-oriented unlearning techniques to some extent compromise the accuracy and effectiveness of the unlearned model $M_u$.
This is because these methods are mainly intended to enhance unlearning efficiency by performing limited retraining on certain subsets of the training set. 
Consequently, 
they cannot provide a strong guarantee like complete retraining does and may result in a partial loss of model performance.
On the other hand, 
these methods lack verification of unlearning results and robustness (privacy protection) testing. 
We believe this is potentially due to the lack of unified design criteria, 
which consequently overlooks the users' demand for verifying whether data has been successfully removed.


Furthermore,
we pay particular attention on the privacy vulnerability of data-oriented unlearning techniques.
Existing research has demonstrated that SISA is vulnerable to member inference attacks \cite{Chen2021when},
since the output distributions of the learned and unlearned models can be used to construct feature vectors to train the attack model.
The differences between two models can reveal the membership information of unlearned data.
It can be seen that the difference in outputs of two models is crucial for adversaries to steal information. 
Specifically,
for data partition methods,
they limit the impact of data on the model through dividing the original training set,
and achieve unlearning by retraining submodel on the subset which contain $D_f$.

However,
the retrained submodels will be reflected in the aggregated confidence vectors,
which results in discrepancy of the outputs before and after unlearning.
This may lead to the disclosure of the unlearned data's information,
like membership (sample-level unlearning). 
When it comes to graph data,
the removal of nodes and edges involve structural information,
so graph unlearning is more likely to generate distinctions.
Meanwhile,
for machine learning models with simple structure like $k$-means \cite{Ginart2019},
it may also be vulnerable to model extraction attacks,
where an adversary can replicate the learned model's behaviors to obtain sensitive information.

For data modification methods,
they involve transforming data or adding noise to erase information,
such as summation form \cite{Cao2015} and noise matrix for class-unlearning \cite{Tarun2021}.
Whether it is data transformation or noise,
there are certain attributes involved that can be inferred through queries,
causing the exposure of these attributes in the differences of outputs between the learned and unlearned models.
Therefore,
data modification methods are more vulnerable to attribute inference attacks, which determines whether the unlearned attribute exist in the original training set.
Table \ref{comparisondata} presents a summary and comparison of these methods.

\begin{table*}[!h]  \scriptsize
	\centering
	\caption{Summary and Comparison of Data-oriented Unlearning Techniques}
	\label{comparisondata}
	\begin{threeparttable}
		\resizebox{\textwidth}{!}{
		\begin{tabular}{cc|cccccc}
			\hline
   \toprule
			\multicolumn{2}{c|}{\makecell[c]{Unlearning \\ Techniques}} & \makecell[c]{Original \\ Models} & \makecell[c]{Request \\ Types} & Accuracy & Effectiveness & Efficiency & Privacy Vulnerability \\ 
   \midrule
			\multirow{5}{*}{\makecell[c]{Data \\ Partition}} & 
			\makecell[c]{Boutoule et al. \cite{Bourtoule2021}} & DNN & \makecell[c]{Batch and \\ stream} & \ding{55} & \ding{55} & \ding{51} & Membership inference attack\\ \cline{2-8}
			
			& \makecell[c]{Ginart et al. \cite{Ginart2019}} & k-Means & \makecell[c]{Sample and \\ stream} & \ding{55} & \ding{55} & \ding{51} & Membership inference attack*\\ \cline{2-8}
			
			& \makecell[c]{Gupta et al. \cite{Gupta2021}} & \makecell[c]{Non-convex \\ models} & \makecell[c]{Adaptive \\ stream} & \ding{55} & \ding{55} & \ding{51} & Membership inference attack*\\ \cline{2-8}
			
			& \makecell[c]{Chen et al. \cite{Chen2022}} & GNN & \makecell[c]{Node and \\ edge} & \ding{55} & \ding{55} & \ding{51} & Membership inference attack*\\ \cline{2-8}
			
			& \makecell[c]{Wang et al. \cite{Wang2023inductive}} & GNN & \makecell[c]{Node, edge \\ and feature} & \ding{55} & \ding{55} & \ding{51} & Membership inference attack*\\ \cline{1-8}
			
			\multirow{3}{*}{\makecell[c]{Data \\ Modification}} & 
			\makecell[c]{Cao et al. \cite{Cao2015}} & \makecell[c]{Statistical query and \\ Bayes models} & \makecell[c]{Sample} & \ding{51} & \ding{51} & \ding{51} & Attribute inference attack*\\ \cline{2-8}
			
			& \makecell[c]{Tarun et al. \cite{Tarun2021}} & DNN & \makecell[c]{Class} & \ding{55} & \ding{55} & \ding{51} & Attribute inference attack*\\ \cline{2-8}
			
			& \makecell[c]{Guo et al. \cite{Guo2022}} & DNN & \makecell[c]{Attribute} & \ding{51} & \ding{51} & \ding{51} &  Attribute inference attack*\\ 
   \bottomrule
		\end{tabular}
	}
	\end{threeparttable}
\end{table*}

\subsection{Model-oriented Techniques}


\subsubsection{Model Reset}

Model reset aims to offset the impact of forgotten data on the model through directly updating the model parameters.
These methods can mainly be categorized into various types,
unlearning via influence function and Fisher information,
but there are also other methods.

\paragraph{Unlearning via Influence Function}

Influence function is used to quantify the impact of sample on model parameters estimation based on the influence theory \cite{Koh2017}.
Guo et al. \cite{Guo2020} proposed the certified removal (CR) mechanism.
Inspired by differential privacy \cite{Wang2020empirical},
they designed a general unlearning framework with strong theoretical guarantee for data removal from trained models.
Unlike differential private learning algorithm,
which ensures that each sample in the training set has negligible impact to the outputted model \cite{Abadi2016},
certified removal only need to unlearn the requested samples.
Specifically,
it can be divided into two steps.
Firstly,
restricting the maximum difference (max-divergence) between models trained on $D$ and $D_r=D\backslash D_f$.
Next,
it utilizes the one-step Newton update \cite{Koh2017} on the model parameters to remove the influence of the deleted sample.
It also applies a loss perturbation approach \cite{Chaudhuri2011} to perturb the residual error,
ensuring the efficiency of certified removal.
Notably,
this mechanism is practical for data removal from
linear/non-linear logistic regressors.
However, 
the Newton update process requiring constructing and inverting the Hessian matrix \cite{Koh2017},
which may cause extremely expensive computational complexity.
This mechanism also does not support removal from models with non-convex losses.

There are other literature expanded and improved the certified removal mechanism.
Izzo et al. \cite{Izzo2021} proposed the first approximate unlearning algorithm for linear and logistic model,
Projective Residual Update (PRU),
which has $\emph{O}(d)$ computational cost.
The basic idea is to calculate the projection of exact parameter update into a low-dimensional space,
namely an approximate parameter update.
This method first calculates the pseudo-inverse matrix of the forgotten data $D_f$,
then using gradient descent to calculate its product with the original model parameters $w_o$
and update the parameters to removal $D_f$.
The PRU achieves a lower computational cost linear in the the feature dimension $O(d)$,
outperforming most the gradient-based methods with superlinear time dependence on the dimension~\cite{Guo2020, giordano2019swiss}.


Similarly,
Neel et al. \cite{Neel2021} extended the certified removal mechanism in the context of data deletion from convex models,
and proposed regularized distributed gradient descent and perturbed distributed gradient descent algorithms.
Its key idea is to publish the unlearned model $M_u$ with added noise, 
optimizing the trade-off between accuracy and efficiency.
Chourasia et al. \cite{Chourasia2023} proposed a data deletion method based on noisy gradient descent.
The model parameters are updated through adding noise after each deletion operation,
to ensure the privacy of unlearned data while reducing computational complexity.

Considering the deletion requests for graph data,
there are also researches investigating graph unlearning based on the theory of certified removal \cite{Guo2020}.
Chien et al. \cite{chien2023efficient} proposed an approximate graph unlearning mechanism,
and the deletion requests include nodes,
edges and node features.
They provided detailed theoretical analysis of approximate graph unlearning for SGC and its generalized PageRank models.
This method not only improves computational efficiency but also effectively defends against member inference attacks.

Wu et al. \cite{Wu2023certified} proposed a certified edge unlearning framework,
which updates the model parameters of the pre-trained GNNs in a single step.
The framework consists of two parts.
First,
it involves adding the perturbation to the loss function,
to hide the gradient residual.
Then,
it estimates the one-shot update on the model's parameters,
trained with the noisy loss function in the first part, based on the influence theory.
This method provides a more efficient data deletion process,
and a certified guarantee.

\paragraph{Unlearning via Fisher Information}

The second type of model reset techniques utilizes the Fisher information \cite{Martens2020} to implement unlearning,
which provides the amount of information about model parameters carried by data.
Golatkar et al. \cite{Golatkar2020} proposed selective forgetting in deep neural networks (DNNs).
It is also a generalized and weaker form of Differential Privacy.
The goal of \emph{selective forgetting} is to minimize the information about a particular
subset $D_f$ selected by the user,
rather than any sample in the training dataset (as defined by differential privacy).
Although this method can maintain good accuracy,
its applicability is low due to the various theoretical limitations.

Golatkar et al. \cite{Golatkar2021} also proposed an ML-Forgetting algorithm based on the mixed-privacy setting.
They divided the training set into two groups, 
core data (not need to be forgotten), 
and user data.
The core weights and user weights corresponding to the two types of data are obtained through standard training and solving a strongly convex quadratic optimization problem.
Based on this assumption, 
the removal of user data can be achieved through quadratic unlearning algorithms,
i.e. minimizing a quadratic loss function.
They derived the amount of remaining information about unlearned data that can be extracted from the weights in a potential attack,
which provides privacy guarantees.
However,
this algorithm compromised the accuracy as it relies on the strongly convex nature of the loss function.

\paragraph{Other Methods}

Information leakage is not only manifested on single data points (i.e. sample-level), 
but features and labels may also contain users' private information.
For example, in a face image dataset,
a deletion request to unlearn a user generally needs to delete all the images of the given user,
and these images have the same label, say, the user's name.
For feature-level unlearning,
Warnecke et al. \cite{Warnecke2021} proposed a data removal algorithm based on influence theory.
The core idea is to evaluate the effect of changing features and labels first and
further to update trained model according to the effect.
They designed two strategies for the close-form update of the model parameters to update the model and remove the influence.
One is the first-order update,
which can be applied to any model with a differentiable loss function.
And the other strategy is the second-order update,
which is only for loss functions with an invertable Hessian matrix.
This method was verified to be theoretically valid by differential privacy and certified removal mechanism,
and provided a significant acceleration for unlearning in experimental analysis.

\subsubsection{Model Modification}

Model modification modifies the model parameters associated with the forgotten data.
These methods usually compute and store all intermediate model parameters during the training process in advance,
and unlearn a data by iteratively updating or replacing model parameters.
Based on different model structures,
model modification techniques fall into three types,
tree-based models unlearning,
GANs unlearning and DNNs unlearning.

\paragraph{Tree-based Models Unlearning}

In tree-based models,
unlearning is usually achieved by computing the parameters of all subtrees and replacing the affected portions.
The random forest is a typical tree-based ensemble model that
utilizes both bagging and feature randomness to generate an uncorrelated forest of decision trees.
To enable the efficient removal of training instances from a random forest model,
Brophy et al.~\cite{Brophy2021} proposed a Data Removal-Enabled Forest (DaRE).
DaRE is an exact unlearning process,
meaning that the resulting unlearned model is identical to the retrained model.
Each tree in a DaRE forest is
trained independently on a copy of the training data.
Thus, given an instance to be unlearned,
only the trees to which the instance belongs need to be retrained.
In addition,
DaRE forests leverage randomness and storing statistics at each node in the tree
to supports efficient model updates.


Wu et al. \cite{Wu2023delta} explored unlearning in Gradient Boosting Decision Trees (GBDT),
and proposed DeltaBoost,
an efficient data removal method.
They used bin-tree to construct a robust histogram,
to ensure that deleting data does not generate new split candidates. 
They also minimized the negative impact on model performance through bagging and gradient quantization.
Bagging technology can make each unlearned data only affects a portion of trees,
and the gradient quantization can be used to reduce the dependency among trees,
which enables unlearning without updating the entire model.
DeltaBoost will provide efficient unlearning process without the loss of accuracy.

\paragraph{DNNs Unlearning}

In model modification,
there are other researchers working on unlearning for Deep Neural Networks (DNNs).
Chundawat et al. \cite{Chundawat2022can} proposed an approximate unlearning method based on student-teacher framework,
which can remove the information about $D_f$ by using a student model and two teacher models (competent and incompetent).
They trained the student model on the original training set,
where the incompetent teacher model provided bad knowledge about $D_f$,
and the competent teacher model provided correct information about $D_r$,
to remove data points and maintain model performance.
Additionally,
they introduced an evaluation metric,
ZRF,
which does not rely on the retrained model.
This unlearning method has generalization in several types of DNNs,
while keeping the accuracy of the model.

Chundawat et al. \cite{Chundawat2022zero} also proposed a zero-shot unlearning approach,
Gated Knowledge Transfer (GRT).
This approach uses pseudo data generated by the generator to train the model based on the weight parameters of the learned model.
A band-pass filter is introduced to ensure that the model only learns information about the retained data $D_r$,
rather than the forgotten data $D_f$.
Moreover,
they designed the Anamnesis Index to measure the quality of unlearning,
which compares the runtime of the unlearned model with it of retraining.
GRT not only have good performance in DNNs unlearning,
but also can effectively defend against membership inference and model inversion attack.

Moreover,
for class-level unlearning in DNNs,
Baumhauer et al. \cite{Baumhauer2022} designed an unlearning algorithm using linear filtration,
which offers both effectiveness and provability for unlearning at the class-level in classification tasks.
Specifically,
they defined the class $\mathcal{F}$ of all classifiers $f \in \mathcal{F}$.
Let $\mathcal{C}$ be a set of classes that wants to be unlearned,
a map
$\mathcal{U}: \mathcal{F} \rightarrow \mathcal{F}_{\neg \mathcal{C}}$,
is said to be an unlearning operation that
unlearns $\mathcal{C}$ from $\mathcal{F}$ if
$\mathcal{U}(A(D))$ and $A_{\neg \mathcal{C}}(D_{{r}_{\neg\mathcal{C}}})$ have the same distribution over $\mathcal{F}_{\neg \mathcal{C}}$.
They further proposed the normalizing filtration unlearning method to achieve class-level unlearning.
Its idea is to proportionally shift the predictions of the unlearned class to the remaining classes.
It turns out that normalizing filtration unlearning ensures that
the probability distribution predicted for the unlearned class after unlearning is approximated to the distribution
predicted by the model retrained without the unlearned class.
This method is also capable of mitigating potential privacy risks arising from model inversion \cite{Veale2018}.

\paragraph{GANs Unlearning}

Generative Adversarial Networks (GANs) applied two neural networks that compete with each other to improve accuracy of predictions \cite{Goodfellow2020, Hirschberg2015}.
For unlearning in GANs,
Kong et al. \cite{Kong2023} proposed a fast approximate data deletion method.
This method is based on a density ratio framework,
using the variational divergence minimization algorithm to train a density ratio estimator $\hat{\rho}_{\epsilon}$ between the original training set $D$ and the remaining training set $D_r$,
to approximate the density ratio $\hat{\rho}$ between the original model $M_o$ and retrained model $M_r$.
Moreover,
they also designed a statistical test mechanism through likelihood ratio and ASC statistic,
to determine whether data has been deleted.
This method has been proved effective in generative models,
which reduce the generation of samples with certain labels,
but it could be difficult to apply to large datasets.

Bae et al. \cite{Bae2023} proposed an unlearning approach,
gradient surgery for pre-trained generative models.
This approach is inspired by multi-task learning \cite{Yu2020gradient},
aiming to offset the influence of target data on the weight parameters through modifying model gradients.
The gradients modification is achieving by projecting gradients onto the normal plane of the retained gradients to regularize the impact among samples,
thereby successfully unlearning data.
It maintains model performance, 
and after unlearning, 
the generated images by the GAN no longer contain samples with specific features.

Moon et al. \cite{Moon2023} proposed a feature-level unlearning framework for generative models,
like GANs and VAEs.
This framework mainly includes three steps.
First,
it collects randomly generated data that contains target feature.
Then,
it identifies  latent representation that represents the target feature in the latent space.
Finally,
it uses this latent representation to implement backpropagation on target feature to fine-tune the generator until it does not produce the target feature.
This unlearning method can remove target features while maintaining the fidelity of model.
Moreover,
the unlearned model is more robust against adversarial attacks.
Sun et al. \cite{Sun2023} also proposed a cascaded unlearning algorithm for pre-trained GANs.
They introduced a substitute mechanism that replaces the data undergoing the unlearning process to maintain the continuity of the latent space,
as well as a fake label to fix the unlearning criterion of the discriminator.
This method implements the unlearning and learning process in a cascaded manner,
which achieves efficient unlearning,
and effectively prevents over-unlearning,
providing a strong privacy guarantee.

\subsubsection{Discussion and Privacy Vulnerability of Model-oriented Techniques}



Most of model-oriented unlearning compromises model accuracy and effectiveness.
Possible reasons for this phenomenon mainly include:
firstly,
the computational complexity of model parameters will affect the trade-off between model performance and deletion efficiency;
secondly,
most of these methods implement approximate unlearning,
which allow for certain difference between the learned and unlearned models;
thirdly,
reset and replacement of parameters may have negative impacts on model predictions.

We also discuss the privacy risks in model-oriented unlearning techniques.
Firstly,
existing study has shown that poisoning attacks not only can reveal privacy information of unlearned data in certified removal mechanism \cite{Marchant2022},
but also can offset the advantages of retraining from scratch,
thereby slowing down the unlearning process.
It means that even without manipulating the original training set,
and limiting the differences between the unlearned and retrained models,
it does not completely eliminate the possibility of attacks and does not enhance the model's robustness.

Secondly,
for model reset methods,
which offset the impact of unlearned model on the model by directly updating model parameters.
Specifically,
these methods generally utilize a loss perturbation technique that hides the gradient residual to perturb the training loss,
which prevents adversaries to a certain extent from extracting information in the unlearned model.
However,
when removing the poisoned data,
the model needs to update parameters to adapt to the remaining training set.
These poisoned samples can trigger backdoors,
which can expand the difference between the unlearned and learned models,
and force model to be retrained to minimize the gap.
It may cause that the information the model unlearned exceeds what it ought to unlearn,
thereby reducing the efficiency of unlearning and destroying the model performance.
Thus,
model reset is vulnerable to poisoning attacks.

Thirdly,
for model modification methods,
which precompute all intermediate parameters,
and then updates or replaces the model to unlearn data.
Due to these methods storing all parameters,
model inversion attacks under white-box settings may be easier to implement.
Such attacks allow adversaries to infer information of the training data from model predictions.
Particular in GANs unlearning \cite{Bae2023},
they can be used to reconstruct the deleted generated image data.
Moreover,
when it comes to unlearning data with specific features generated by generator in GANs,
it is also possible to steal information through attribute inference attacks.

Finally,
it is worth noting that there are several methods \cite{Baumhauer2022, Moon2023, Chundawat2022can, Chien2023} that meet the requirements of verifiability and robustness.
These methods not only provide a verification mechanism for the forgotten data,
but also demonstrate their robustness against membership inference attacks and model inversion attacks.
This highlights the significance of secondary privacy leak problem in unlearning.
Meanwhile,
several studies have provide theoretical guarantees,
demonstrating that the model weights no longer contain information of unlearned data through theoretical analysis.
The summary and comparison of model-oriented unlearning methods are shown in Table \ref{comparisonmodel}.

\begin{table*}[!h] \scriptsize
	\centering
	\caption{Summary and Comparison of Model-oriented Unlearning Techniques}
	\label{comparisonmodel}
	\begin{threeparttable}
		\resizebox{\textwidth}{!}{
		\begin{tabular}{cc|cccccc}
			\hline
   \toprule
			\multicolumn{2}{c|}{\makecell[c]{Unlearning \\ Techniques}} & \makecell[c]{Original \\ Models} & \makecell[c]{Request \\ Types} & Accuracy & Effectiveness & Efficiency & Vulnerability \\ 
   \midrule
			\multirow{9}{*}{\makecell[c]{Model \\ Reset}} & 
			\makecell[c]{Guo et al. \cite{Guo2020}} & \makecell[c]{Linear \\ models} & \makecell[c]{Sample and \\ batch} & \ding{55} & \ding{55} & \ding{51} & Poisoning attack\\ \cline{2-8}
			
			& \makecell[c]{Golatkar et al. \cite{Golatkar2020}} & DNN & \makecell[c]{Sample and \\ class} & \ding{55} & \ding{55} & \ding{51} & Poisoning attack* \\ \cline{2-8}
			
			& \makecell[c]{Golatkar et al. \cite{Golatkar2021}} & DNN & \makecell[c]{Sample and \\ class} & \ding{55} & \ding{55} & \ding{51} & \ding{51} \\ \cline{2-8}
			
			& \makecell[c]{Izzo et al. \cite{Izzo2021}} & \makecell[c]{Linear and \\ logistic models} & \makecell[c]{Sample and \\ batch} & \ding{55} & \ding{55} & \ding{51} & Poisoning attack*\\ \cline{2-8}
			
			& \makecell[c]{Neel et al. \cite{Neel2021}} & \makecell[c]{Convex \\ models} & \makecell[c]{Stream} & \ding{55} & \ding{55} & \ding{51} & Poisoning attack*\\ \cline{2-8}
			
			& \makecell[c]{Chouasia et al. \cite{Chourasia2023}} & \makecell[c]{Convex and \\non-convex models} &  \makecell[c]{Adaptive \\stream} & \ding{55} & \ding{55} & \ding{51} & Poisoning attack*\\ \cline{2-8}
			
			& \makecell[c]{Chien et al. \cite{Chien2023}} & GNN & \makecell[c]{Node, edge and \\ feature}  & \ding{55} & \ding{55} & \ding{51} & \ding{51}\\ \cline{2-8}
			
			& \makecell[c]{Wu et al. \cite{Wu2023certified}} & GNN & \makecell[c]{Edge}  & \ding{55} & \ding{55} & \ding{51} & Poisoning attack*\\ \cline{2-8}
			
			& \makecell[c]{Warnecke et al. \cite{Warnecke2021}} & \makecell[c]{Convex and \\ non-convex models} & \makecell[c]{Feature and \\ label} & \ding{55} & \ding{55} & \ding{51} & Attribute inference attack*\\ \cline{1-8}
			
			\multirow{9}{*}{\makecell[c]{Model \\ Modification}} & 
			\makecell[c]{Brophy et al. \cite{Brophy2021}} & \makecell[c]{Random \\ forest} & Batch & \ding{51} & \ding{51} & \ding{51} & Model inversion attack*\\ \cline{2-8}
			
			& \makecell[c]{Wu et al. \cite{Wu2023delta}} & \makecell[c]{Decision \\ tree} & Batch & \ding{51} & \ding{51} & \ding{51} & Model inversion attack*\\ \cline{2-8}
			
			& \makecell[c]{Kong et al. \cite{Kong2023}} & GAN & Sample & \ding{51} & \ding{51} & \ding{51} & Model inversion attack*\\ \cline{2-8}
			
			& \makecell[c]{Bae et al. \cite{Chen2021gan}} & GAN & \makecell[c]{Class and \\ feature} & \ding{51} & \ding{51} & \ding{51} & Model inversion attack* \\ \cline{2-8}
			
			& \makecell[c]{Moon et al. \cite{Moon2023}} & GAN & Feature & \ding{55} & \ding{55} & \ding{51} & \ding{51} \\ \cline{2-8}

            & \makecell[c]{Sun et al. \cite{Sun2023}} & GAN & \makecell[c]{Sample and \\class} & \ding{55} & \ding{51} & \ding{51} & \ding{51} \\ \cline{2-8}
			
			& \makecell[c]{Chundawat et al. \cite{Chundawat2022can}} & DNN & \makecell[c]{Sample and \\ class} & \ding{55} & \ding{55} & \ding{51} & \ding{51}\\ \cline{2-8}
			
			& \makecell[c]{Chundawat et al. \cite{Chundawat2022zero}} & DNN & \makecell[c]{Sample and \\ class} & \ding{55} & \ding{55} & \ding{51} & \ding{51}\\ \cline{2-8}
			
			& \makecell[c]{Baumhauer et al. \cite{Baumhauer2022}} & DNN & Class & \ding{55} & \ding{55} & \ding{51} & \ding{51} \\ 
   \bottomrule
		\end{tabular}
	}
	\end{threeparttable}
\end{table*}

\section{Privacy Threats of Machine Unlearning}\label{privacythreat}


\subsection{Information-stealing Attacks}

Information-stealing attacks refer to conducting attacks after using the differences between the learned model and unlearned model
to obtain sensitive information.
This type of privacy threat is mainly caused by the two versions of the machine learning models generated before and after unlearning.
Possible methods include membership inference attack, 
attribute inference attack, 
model inversion attack, 
and model extraction attack.

\subsubsection{Membership Inference Attack}\label{privacymia}

\begin{figure}[h]
	\centering
	\includegraphics[scale=0.85]{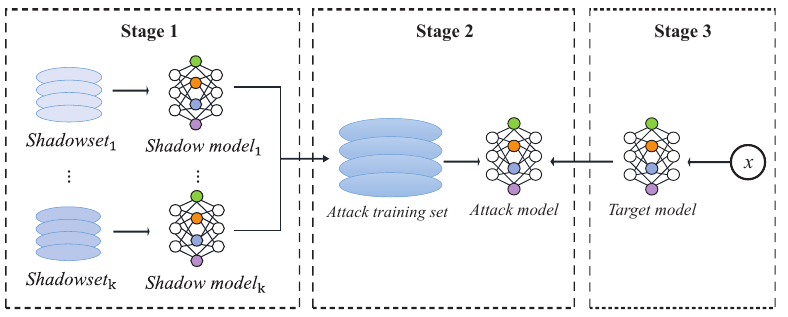}
	\caption{Membership Inference Attack}
	\label{mia}
\end{figure}

Membership Inference Attack (MIA) \cite{Shokri2017},
aims to determine whether a target sample belongs to a model's training set.
The core idea of MIA is to construct several shadow models that generate datasets to train the attack model based on the differences of the machine learning model on the training data and the first encountered data. Fig. \ref{mia} illustrates the detailed framework of Membership Inference Attack,
in which the process consists of three main stages.

\begin{itemize}
	\item \textbf{Stage 1: Train shadow model.}
	The attacker constructs $k$ shadow models,
	denoted as $\mathcal M_{shadow,i}$.
	Each shadow model is trained on a disjoint subset of the training set $D$, 
	where each subset $D_{train shadow,i}$ contains data records and their corresponding labels. 
	The shadow models then output the prediction vectors, $y$, for the data samples.
	\item \textbf{Stage 2: Train attack model.}
	For the output $y$ of the shadow models on the training data, 
	the record $(label, y, in)$ is added to the training set $D_{train attack}$ of the attack model. 
	Similarly, 
	for the output of the shadow models on the test data, 
	the record $(label, y, out)$ is added to $D_{train attack}$. 
	Based on the label values, 
	$D_{train attack}$ is divided into several partitions associated with different class labels,
	and a binary classifier is trained on each partition.
	\item \textbf{Stage 3: Membership inference.}
	Given a record $(x, label)$, 
	the data sample $x$ is input to the target model to obtain the prediction vector $y$. 
	Then, 
	based on the $label$,
	$y$ is input to the corresponding classifier to infer whether the sample $x$ belongs to the training set of the target model or not.
\end{itemize}

MIA allows for the construction of the attack training set 
and the training of the attack model by designing several shadow models.
This enables the inference of whether a data sample is a member of the training set of the target model.

Since there are two models generated before and after machine unlearning,
and the differences between them may contain private information about the forgotten data, 
resulting in a privacy disclosure, 
Chen et al. \cite{Chen2021when} firstly investigated this problem caused by unlearning and proposed a novel attack method based on the idea of membership inference attack.
The basic idea of this attack is to determine whether the forgotten data $D_f$ is a member of the training set $D$ of the original model ${M}_o$ by using the differences between the two models generated before and after unlearning. 
Unlike classical membership inference \cite{Shokri2017}, 
which only utilizes the output posterior of the target model, 
this attack method leverages the comprehensive information of the learned model ${M}_o$ and the unlearned model ${M}_u$. 

\begin{figure*}[h]
	\centering
	\includegraphics[scale=0.6]{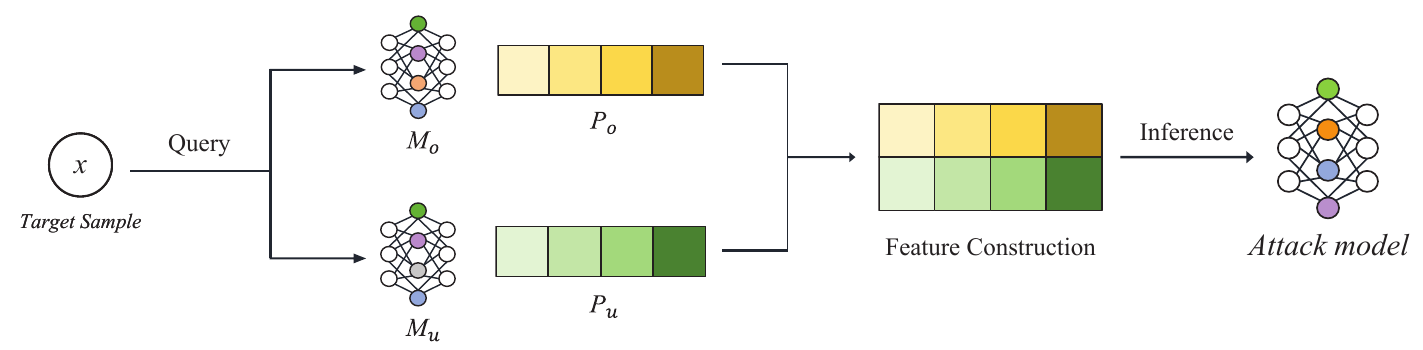}
	\caption{Membership Inference Attack In Unlearning}
	\label{when}
\end{figure*}

The workflow of membership inference attack in unlearning is shown in Fig. \ref{when}.
This attack method is consists of three phases.
In the first phase, 
the adversary trains an attack model using shadow models, 
which include the original learned model and several shadow unlearned models. 
The second phase is generating posteriors. 
Given a target sample $x$ that needs to be forgotten,
they use the unlearned model $M_u$, 
obtained through retraining, 
along with the learned model $M_o$, 
to acquire the corresponding posteriors $P_o$ and $P_u$ for a target sample $x$ that needs to be forgotten. 
These two posteriors are then combined to construct a feature vector $P^{'}$ for $x$. 
Finally, 
the vector $P^{'}$ is inputted into the attack model, 
a binary classifier, 
to determine whether $x$ is a member of the training set $D$ of the learned model $M_o$. 
This method can quantify the difference between $M_o$ and $M_u$ and provide evidence of secondary privacy disclosure in machine unlearning.

Furthermore,
Gao et al. \cite{Gao2022} formalized concepts related to deletion inference and deletion reconstruction in the context of machine unlearning. 
Deletion inference refers to the goal of distinguishing between a deleted data sample $x$ from the training set $D$ of a machine learning model and another sample $x^{'}$ that is not deleted. 
Inference attacks based on membership inference have been shown to be effective in regression and classification tasks.
Deletion reconstruction, 
on the other hand, 
focuses on reconstructing the deleted sample or deletion label. 
The purpose of deletion sample reconstruction attack is to acquire the features of the deleted sample $x$. 
And in the case of deletion label reconstruction, 
deleting a data sample with label $c_1$ may reduce the probability that the unlearned model $M_u$ outputs label $c_1$. 
Thus, 
deletion reconstruction attacks aim to obtain the label information of the deleted sample, 
potentially leading to privacy leakage in unlearning.

\subsubsection{Attribute Inference Attack}

Attribute inference attack aims to obtain individual sensitive attributes based on collection and analysis of observable features.
This attack is mainly in the context of white-box attacks,
involving inferring hidden or incomplete attribute data by utilizing publicly available attributes and model structures \cite{Liu2022ml}.
The basic process of attribute inference attack is normally consisted of the following steps \cite{Ganju2018}.


\begin{itemize}
	\item \textbf{Stage 1: Construct attribute dataset.}
    The attacker selects $k$ (at least two) attributes about the training set $D$ of the target model,
    denoted as $A_k$ ($k=1, 2, \ldots, n$),
    for example,
    $A_1$ and $A_2$.
    Then constructs attribute datasets $D_{trainshadow,1}$ and $D_{trainshadow,2}$ that fulfill the respective attribute.
	\item \textbf{Stage 2: Train shadow model.}	
    The attacker trains several shadow models $M_{shadow,i}$ on each attribute dataset $D_{trainshadow,1}$ and $D_{trainshadow,2}$,
    which have the same architecture as the target model.    
	\item \textbf{Stage 3: Train attack model and inference.}
    After training shadow models,
    using resulting representations as the training set $D_{trainattack}$ to train the attack model $M_{attack}$.
    Then,
    the attacker feed it the representation of target attribute,
    to predict the existence of the target attribute from the target model.
\end{itemize}


The trained attack model $M_{attack}$ enables the attacker to infer the sensitive attributes to gain a deeper understanding of the training data.
In general,
attribute inference attacks can be implemented when statistical characteristics can be inferred from outputs or when attributes can be distinguished. 
Thus,
class-level and feature-level unlearning would be vulnerable to attribute inference attacks.

Stock et al. \cite{Stock2023} investigated attribute inference attacks in feature-level unlearning under a white-box setting.
Like traditional attribute inference attack,
this attack for unlearning targets attribution information about $D_f$,
namely inferring whether there exists a certain attribute in the original training set $D$.
They utilized the attack method in \cite{Ganju2018},
considering $M_o$ as the target model,
to construct shadow datasets $D_{shadow,i}$ which are contained the unlearned attribute.
Then,
training the corresponding shadow models $M_{shadow,i}$ on each shadow dataset,
and the obtained representations is used to train $M_{attack}$.
The representations of $M_u$ are used as the input of $M_{shadow,i}$ to determine whether the unlearned data's attribute information is present in $D$.
It can be seen that attribute inference attacks could be implemented in feature-level or class-level unlearning.
However,
based on the results of \cite{Chen2021when},
whether the difference in representations of an attribute in $M_o$ and $M_u$ can extract the privacy information of $D_f$ still needs to be studied.


\subsubsection{Model Inversion Attack}\label{privacyinversion}

Model inversion attack
refers to an attack method that uses a machine learning model's output results to recover or infer the model's input data \cite{Fredrikson2015}. 
The attacker can reconstruct the original input data through using the information from the model's output with reverse inference techniques or reverse modeling methods.
Image classifiers are particularly vulnerable to this type of attack \cite{Liu2021when}. 
The core idea of model inversion attack is to construct a reverse mapping based on the model's output in order to complete the target vector and recover the input data,
involving three specific stages.

Suppose the target model is $M_{target}: \mathcal X \rightarrow \mathcal Y$, 
and its input-output pairs are denote as $x_i$ and $y_i$. 
The goal of the attacker is to conduct reverse-engineering to find $M_{target}^{-1}$ based on the learned machine learning model,
to recover the input $x_i$ based on the output $y_i$.

\begin{itemize}
	\item \textbf{Stage 1: Construct attack training set.}
	The attacker collects the input vectors $x={x_1, x_2, \ldots, x_n}$ and their corresponding output vectors $y={y_1, y_2, \ldots, y_n}$ of the target model $M_{target}$ through queries,
	which serves as the training set for the attack (reverse) model.
	\item \textbf{Stage 2: Train reverse model.}
	Constructing a reverse model $M_{target}^{-1}$,
	aiming to infer input vectors $x$ from output vectors $y$,
	and training $M_{target}^{-1}$ based on attack training set and gradient in target model $M_{target}$.
	\item \textbf{Stage 3: Inversion inference.}
	Given the output $y_t$ of a specific sample $x_t$ from the target model $M_{target}$, 
	the attacker inputs $y_t$ into a trained reverse model allows for the recovery of relevant information about that sample.
\end{itemize}

Overall,
model inversion attacks can use the gradient information in a learned model to search for the inverse model, 
enabling to obtain the features of all classes. 
This indicates that such attacks are likely to leak information about distinguishable classes \cite{Liu2021when}.
Existing research \cite{Baumhauer2022} had demonstrated that the unlearned model generated by class-level unlearning is vulnerable to model inversion attacks.
Based on the result of a correlation between input and output space,
they used gradient ascend on the input space to reconstruct the unlearned classes' image information.
Similarly,
Graves et al. \cite{Graves2021} proposed a slightly modified version of model inversion attack in \cite{Fredrikson2015} against class-level unlearning,
which serves as an evaluation strategy.
Specifically,
the purpose of this attack is to recover the images of $D_f$.
In the reverse engineering process in Stage 2 mentioned above,
they firstly labeled the feature vector with the label of the target class,
and obtained the corresponding loss gradient through a forward pass.
Secondly,
each feature is shifted in the direction of the gradient,
and iteratively making it closer to the target class.
Finally,
a PROCESS function is applied every $g$ gradient descent steps ($g \in [500, 1000]$) to improve the clarity of the reconstructed image. 
The experimental results had proved the effectiveness of this attack method in reconstructing the unlearned data in class-level unlearning.
Furthermore, 
we believe that gradient-based unlearning methods may be vulnerable to model inversion attacks under white-box settings, 
as gradients can provide more loss information.



\subsubsection{Model Extraction Attack}

Model extraction attack refers to extracting non-public information,
like model parameters,
from a black-box machine learning model.
The goal is to construct a model to clone the target model,
as first proposed by Tramer et al. \cite{Tramer2016}.
Specifically,
they designed a model extraction method based on prediction APIs,
involving the following steps.

\begin{itemize}
	\item \textbf{Stage 1: Construct attack training set.}
	The attacker collects a set of input-output pairs $(x_i, y_i)$ by querying the prediction API,
	which serves as the attack training set $D_{attack} = {(x_1, y_1), (x_2, y_2), \ldots, (x_n, y_n)}$.
	\item \textbf{Stage 2: Train substitute model.}
	Using $D_{attack}$ to train a substitute model $M_{substitute}$,
	which is a  simplified model that approximates the behavior of the target model $M_{target}$.
	\item \textbf{Stage 3: Refine substitute model.}
	The attacker iteratively refines the substitute model based on query results form the  prediction API and compares the outputs with those generated by the substitute model $M_{substitute}$.
	The goal of this iterative process is to minimize the difference between predictions of $M_{substitute}$ and $M_{target}$,
	which is normally quantified by test error and uniform error (the fraction of the full feature space on which two models disagree).
\end{itemize}


This attack can use sufficient queries via prediction APIs to generate a substitute model to simulated the target model.
Jang et al. \cite{Jang2023} implemented model extraction attacks on LLMs unlearning based on the method in \cite{Carlini2021}.
Unlike common extraction attacks,
the purpose of this method is to study the ability of LLMs to memorize data,
instead of create an attack that can target specific users.
It first selects the unlearned token sequence as the target sequence,
and then extracts prior information about the target token sequence through the unlearned LMs,
which involves calculating memorization capability of the model for the target sequence.
However,
model extraction attacks that focus on LLMs unlearning are generally serves as a specific evaluation metric,
which is difficult to apply to mainstream DNNs.
Since the unlearned model generated by the unlearning process leads to changes in model's structure and parameters,
how to utilize these information to effectively implement model extraction attacks in unlearning remains a research gap in this field.


\subsection{Model-breaking Attacks}

Model-breaking attacks refer to methods used by attackers to compromise the integrity of the models generated through the unlearning process.
It is mainly reflected in ensuring that the training process of unlearning and the prediction process of unlearned models are undisturbed, 
such as data deletion efficiency and prediction accuracy.
In this paper, 
the goal of model-breaking attacks is to utilize the vulnerability caused by data deletion operations and the gradient update process, 
thereby reducing model utility and slowing down unlearning efficiency. 
Backdoor attack and poisoning attack are the primary methods to achieve the above goal.

\subsubsection{Backdoor Attack}

Backdoor attack refers to an attack that manipulates the prediction behavior of a model through implanting backdoors,
making it produce specific outputs for certain inputs.
This attack aims to disrupt the model's integrity,
which means that the attacker causes the model to learn specific content,
to maintain high precision for the normal samples while output preset labels for target inputs \cite{Zhou2023}.

Specifically,
the basic idea of backdoor attack is to modify a set of $D$ into poisoned samples,
denoted as $D_b$,
with the specific label $y_b$.
Then,
$D_b$ and normal samples are used together to train a machine learning model that contains hidden backdoor triggers.
When predicting normal samples,
correct results can still be acquired,
while the model will output according to the target class $y_b$ for $D_b$ with triggers.

\begin{figure}[h]
	\centering
	\includegraphics[scale=1]{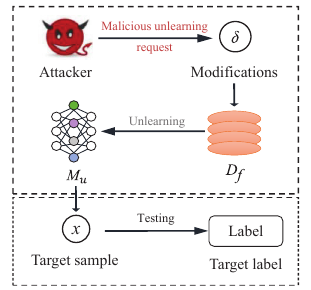}
	\caption{Malicious Unlearning Attack}
	\label{backdoor}
\end{figure}

For unlearning,
the purpose of backdoor attacks is to destroy the unlearned model's integrity by injecting backdoors to mislead prediction results.
Qian et al. \cite{Wei2023} proposed a malicious attack method against DNNs unlearning,
aiming to compromise the unlearned model's prediction capability.
Based on the idea of backdoor attacks,
this method modifies the forgotten data to implant backdoors,
and triggers backdoors during the unlearning process to obtain the unlearned model,
which makes the target test samples' outputs to be designated as the target label.
Fig. \ref{backdoor} illustrates the process of malicious unlearning attack.

Specifically,
the attacker firstly pretends to be a normal user,
and generates a malicious unlearning request
to make the corresponding $\delta$ to modify $D_f$,
i.e. $x_b = x - \delta$ ($x \in D_f$).
Then,
obtaining $M_u$ through the unlearning operation.
Finally,
the attacker uses the target test samples as input of $M_u$ to trigger the backdoor to mislead $M_u$ into outputting specific label $y_b$.
Overall,
this attack triggers the preset backdoor through unlearning,
which causes the unlearned model to misclassify the target test samples.
It has demonstrated the effectiveness of this method in common unlearning techniques,
including exact, like SISA,
and approximate data deletion.

Hu et al. \cite{Hu2023} also proposed a malicious poisoning attack to diminish the prediction capability of the unlearned model,
also called as over-unlearning.
This attack method can makes prediction accuracy of the unlearned model decline by modifying the forgotten data,
even for a specific category.
It first submits data samples to query the model and obtain the corresponding probability vectors,
and then,
based on $D_f$ and query results,
leverages the black-box CW adversarial attack \cite{Carlini2017towards} to construct a perturbed version of $D_f$ for unlearning,
i.e. $x_b = x + \delta$
where $x \in D_f$,
$\delta$ is the perturbation.
Therefore,
this attack method can blend additional samples from different tasks into the original unlearned data.
When unlearning the modified data,
the backdoor will be activated to make the unlearned model output wrong labels,
thereby destroying model's utility.

According to the framework of unlearning,
Zhang et al. \cite{Zhang2023backdoor} proposed a novel backdoor attack through machine unlearning,
called \emph{BAMU}.
This attack introduces carefully designed data containing poison and mitigation samples to implant backdoors.
Then,
the attacker post unlearning requests for removing mitigation samples,
gradually activating the hidden backdoor,
which compromise the model performance.
This backdoor attack can obtain a good learned model,
but once the backdoor is triggered trough unlearning,
the unlearned model with poor utility is obtained.
Similarly,
Di et al. \cite{Di2022} designed a camouflaged data poisoning attack,
making the resulting unlearned model generate misclassifications by polluting the training set with poison and camouflage sets.

\begin{table*}[!h] \scriptsize
	\centering
	\caption{Summary and Comparison of Different Evaluation Metrics}
	\label{comparisonmetric}
	\begin{threeparttable}
		\begin{tabular}{c|c|c|c|c|c}
			\hline
   \toprule
			\multicolumn{3}{c|}{Evaluation Metrics} &
			Principle & Advantages & Disadvantages \\ 
   \midrule
			\multirow{3}{*}{\makecell[c]{Attack-oriented\\ Metrics}} & 
			\multirow{3}{*}{\makecell[c]{Verifiability \\Robustness}} & \makecell[c]{Membership inference\\ Attack} & \makecell[l]{Membership \\information \\ of $D_f$} & \makecell[l]{Intuitive to \\understand} & \multirow{3}{*}{\makecell[l]{Only applicable \\to specific \\unlearning methods}} \\ \cline{3-5}
			& & \makecell[c]{Model inversion \\attack} & \makecell[l]{Degree of \\ reconstructing $D_f$ \\(i.e. image} & \makecell[l]{Good verification \\ability} \\ \cline{3-5}
			& & \makecell[c]{Model backdoor\\ attack} & \makecell[l]{Accuracy of \\attacking $M_u$} & \makecell[l]{Good verification \\ability} \\ 
   \bottomrule
		\end{tabular}
	\end{threeparttable}
\end{table*}

\subsubsection{Poisoning Attack}

Poisoning attack is a deliberate act by the attacker to manipulate the data distribution and reduce the availability of data through consciously putting the wrong or offset data into the training set.
Its goal is to change the behavior of the model and influence its availability \cite{Biggio2012}.
The basic principle of poisoning attack is to poison the original training data to alter the training process,
leading to greatly degraded model utility,
such as a decrease in computational efficiency or prediction accuracy \cite{He2022towards}.
This attack is mainly accomplished through manipulating the training data,
for example,
modifying or poisoning data to disrupt the entire system.



In machine unlearning,
poisoning attacks can be implemented through polluting the partial original training set. 
The goal of this attack is to destroy the unlearned model's availability.
Marchant et al. \cite{Marchant2022} proposed a poisoning attack method against machine unlearning.
Based on the approximation error in certified removal mechanism,
Marchant et al. designed an attack method called slow-down attack,
aiming to offset the computational advantages of unlearning compared with retraining.
The core idea of this method is to increase the error of approximate unlearning update by poisoning the forgotten data,
to trigger retraining more frequently,
thereby reducing the promised efficiency of unlearning.
Specifically,
this attack is formulated as a data poisoning problem, where given a training set $D$ and a learning algorithm $\mathcal{A}$ to train the learned model $M_o$.
Suppose the attacker can poison a subset of samples in $D$, 
denoted as $D_p$, 
while the remaining clean samples are $D_c$. 
When deleting poisoned data $D_p$,
the approximation error exceed the allowed threshold ($\beta > \beta_{trigger}$),
triggering retraining to increase computational cost.
To measure the computational cost $\mathcal{C}(M_o, D_f)$ of removing the forgotten data $D_f$ from the learned model $M_o$ using the unlearning algorithm, 
they introduced a function $\mathcal{C}$. 
For the attacker,
its goal is to maximize $\mathcal{C}(M_o, D_p)$.


The experimental results indicated that this poisoning attack can significantly impede the process of unlearning, 
which requires further investigation. 

\subsection{Attack-oriented Metrics}\label{attackmetrics}

The purpose of machine unlearning is to avoid privacy leaks in machine learning models. 
Privacy attack methods can not only directly evaluate the effectiveness of unlearning operations,
but also provide support for the verifiability and robustness of unlearning,
like membership inference attack, model inversion attack, and model backdoor attack.
While attack-oriented metrics can directly reflect the verifiability and robustness of unlearning techniques,
their applicability requires further research.
Table \ref{comparisonmetric} illustrates a summary of evaluations for unlearning.

\subsubsection{Membership Inference Attack Metrics}

In addition to a privacy attack, MIA can be used to estimate if a sample has been unlearned successfully or not.
The accuracy of MIA represents the probability of membership,
which indicates that confidence that an attack model can identify the forgotten data as a member or non-member. 
Therefore, 
MIA can effectively verify the performance of data removal \cite{Graves2021}.
Chen et al. \cite{Chen2021gan} designed an indicator called the False Negative Rate (FNR), which estimates the ratio of False Negative out of the inferring samples.
The definition is given by Eq. \ref{FNR}.

\begin{equation}\label{FNR}
FNR=\frac{FN}{TP+FN}
\end{equation}

where $TP$ represents MIA inferring the training data as a member of $D$, 
while $FN$ represents MIA inferring the training data as a non-member of $D$,
meaning that the forgotten data would behave like the non-member of training set for the model.
A larger $FNR$ indicates a better performance of unlearning. 
A similar indicator has been proposed by Ma et al. \cite{Ma2022}.

\subsubsection{Model Inversion Attack Metrics}

Model inversion attack exploits the confidence scores of the learned model to obtain data information and replicate the target sample \cite{Fredrikson2015}. 
Therefore, 
the unlearned model following inversion should not contain information about the forgotten data. 
This attack method is commonly used to verify unlearning, 
particularly in image applications \cite{Baumhauer2022}.

\subsubsection{Model Backdoor Attack Metrics}

In the process of backdoor attack,
the adversary inserts hidden triggers into the training process of the target model by adding training samples.
Based on this,
Sommer et al. \cite{Sommer2020} proposed a backdoor verification scheme from the perspective of users,
enabling high-confidence evaluation.
Similarily,
Chen et al. \cite{Chen2021lightweight} provided the definition of backdoor attack in the evaluation of unlearning.
Given a backdoor attack algorithm $\mathcal B \left(\cdot \right)$,
along with the unlearned model $M_u$,
the clean data $(x_i, y_i) \in D$ and the poisoned data $(x_b, y_b) \in D_f$,
if $\mathcal B(M_u(x_b)) = y_i$,
it indicates successful unlearning
Otherwise,
if $\mathcal B(M_u(x_b)) = y_b$,
unlearning fails.



\subsection{Possible Defense Methods for Unlearning Attacks}\label{possibledefense}

\begin{table*}[!h] 
	\centering
	\caption{Overview of Possible Defense Methods}
	\label{defense}
	\begin{threeparttable}
		\resizebox{\textwidth}{!}{
		\begin{tabular}{c|cc}
			\hline
   \toprule
			Attack types & Possible defense & Basic idea \\ 
   \midrule
			\multirow{4}{*}{\makecell[c]{Membership inference attack \\ Attribute inference attack \\ Model inversion attack \\ Model extraction attack}} & Differential Privacy \cite{Jayaraman2019} & \makecell[l]{Updates the model parameters to \\achieve differential privacy} \\ \cline{2-3}
			& Homomorphic Encryption \cite{Jiang2018, Bai2021} & \makecell[l]{Encrypts the original training data} \\ \cline{2-3}
			& Confidence score masking \cite{Hu2022, Jia2019} & \makecell[l]{Hides the true confidence scores} \\ \cline{2-3}
			& Temperature scaling \cite{Guo2017on} & \makecell[l]{Smooths the probability distribution} \\  \hline		
			\multirow{2}{*}{\makecell[c]{Poisoning attack \\Backdoor attack}} & Data preprocessing \cite{Borgnia2021} & \makecell[l]{Cleans or enhance the original \\training data} \\ \cline{2-3}
			& Robust training \cite{Jagielski2018, Jia2021}& \makecell[l]{Adjusts the training algorithm to \\improvie the robustness} \\ 
   \bottomrule
		\end{tabular}
	}
	\end{threeparttable}
\end{table*}

\subsubsection{Defense for Information-stealing Attacks}

\paragraph{Differential Privacy (DP)}

DP is a cryptographic scheme,
which provides a guarantee that each data point in the training set has limited influence on the model outputs.
It can maximize the accuracy of data queries,
while minimizing the probability of identifying their records when querying from a statistical database.
Previous research has demonstrated that DP can effectively prevent inference attacks and inversion attacks \cite{Jayaraman2019, zhu2020}.

\paragraph{Homomorphic Encryption (HE)}

Homomorphic encryption is a form of encryption that allows computations to be performed on encrypted data without first having to decrypt it \cite{Jiang2018}.
Previous studies have shown that through utilizing homomorphic encryption technology to encrypt the update parameters of local clients,
the server can perform computational operations based on the encrypted data,
which protects local data's privacy and effectively defends against MIAs \cite{Bai2021}.

\paragraph{Confidence Score Masking}

Confidence score masking refers to mitigating the effectiveness of attacks by hiding the true confidence scores outputted by the target model \cite{Hu2022}.
It can be implemented in two ways.
One is providing incomplete prediction vectors,
such as only publishing the top-k confidence scores \cite{Shokri2017}.
Another one is adding noise to the outputted prediction vectors.
For example,
Jia et al. \cite{Jia2019} proposed a defense scheme called MemGuard,
which can reduce the information-stealing attacks to a random guess level.

\paragraph{Temperature Scaling}

Temperature scaling is a technology for tuning the sharpness of a model distribution,
which aims to modify the confidence scores to increase model uncertainty \cite{Guo2017on}.
Through adjusting the temperature parameter,
a higher value is introduced to make probability distribution of the model become smoother and more balanced,
increasing the uncertainty for attacks.

\subsubsection{Defense for Model-breaking Attacks}

Backdoor attack and poisoning attack can be mitigated through taking precautions,
mainly including data preprocessing and robust training \cite{Wang2022threats}.

\paragraph{Data Preprocessing}

Data preprocessing is the method of cleaning or enhancing original training data to avoid data pollution.
It has been proven that data augmentations are an effective strategy to defend against poisoning attacks.
For example,
Borgnia et al. \cite{Borgnia2021} utilized convex combinations of data and labels to improve model robustness.

\paragraph{Robust Training}

Robust training aims to adjusting training algorithms for a more robust training process.
For example,
the TRIM defense algorithm, proposed by Jagielski et al. \cite{Jagielski2018},
enables the identification of poisoning data samples and the training of a more robust model.
Ensemble learning method can also defend against poisoning attacks,
for example,
training several submodels on random subsets of the training set and using a majority vote to predict labels \cite{Jia2021}. Table \ref{defense} provides an overview of possible defense methods.

\subsection{Summary}

\begin{table*}[!h] \scriptsize
	\centering
	\caption{Summary and Comparison of Various Privacy Threats}
	\label{comparisonattack}
	\begin{threeparttable}
		\resizebox{\textwidth}{!}{
		\begin{tabular}{cc|cccc}
			\hline
   \toprule
			\multicolumn{2}{c|}{\makecell[c]{Machine Unlearning \\ Privacy Threats}} &
			Goals & \makecell[c]{Target \\ Techniques} & Vulnerabilities & \makecell[c]{Possible \\ defense}\\ 
   \midrule
			\multirow{4}{*}{\makecell[l]{Information\\-stealing \\ Attacks}} & 
			\makecell[l]{Membership \\ Inference \\ Attack \cite{Chen2021when}} & \makecell[l]{Detect whether $D_f$ \\ is a member of $D$} & \makecell[l]{Data \\partition} & \makecell[l]{The difference in \\ outputs between \\ $M_o$ and $M_u$} &  \multirow{4}{*}{\makecell[l]{Differential privacy \\ Homomorphic encryption \\ Confidence score masking \\ Temperature scaling}}\\ \cline{2-5}
			& \makecell[l]{Attribute\\ Inference \\ Attack \cite{Stock2023}} & \makecell[l]{Obtain sensitive \\attributes of $D_f$} & \makecell[l]{Data \\modification*} & \makecell[l]{The difference in \\feature representations\\between $M_o$ and $M_u$\\(e.g. attribute-level unlearning)*} \\ \cline{2-5}
			& \makecell[l]{Model\\ Inversion \\ Attack \cite{Baumhauer2022, Graves2021}} & \makecell[l]{Recover the \\information of $D_f$ \\(e.g. image)} & \makecell[l]{Model \\modification} & \makecell[l]{The distinguishable \\ classes in $M_u$ \\(e.g. class-level unlearning)\\ or gradient sharing*} \\ \cline{2-5}
			& \makecell[l]{Model \\Extraction \\ Attack \cite{Jang2023}} & \makecell[l]{Clone $M_o$ to acquire \\information of $D_f$} & - & - \\ \cline{1-5}
			\multirow{2}{*}{\makecell[l]{Model\\-breaking \\ Attacks}} & \makecell[l]{Backdoor \\ Attack \cite{Hu2023, Wei2023, Zhang2023backdoor}} & \makecell[l]{Compromise $M_u$'s integrity\\ and mislead prediction} & \makecell[l]{All unlearning \\techniques*} & \makecell[l]{Data deletion operations} & \multirow{4}{*}{\makecell[l]{Data preprocessing \\Robust training}}\\ \cline{2-5}
            & \makecell[l]{Poisoning \\ Attack \cite{Marchant2022}} & \makecell[l]{Destroy $M_u$'s availability\\ and slow don \\unlearning} & \makecell[l]{Model \\reset} & \makecell[l]{Approximation errors \\ and gradient update\\ in approximate unlearning} \\  
   \bottomrule
		\end{tabular}
	}
	\end{threeparttable}
\end{table*}

The main privacy threat to machine unlearning is stealing the relevant information about the forgotten data $D_f$, as there are differences between the learned model $M_o$ and unlearned model $M_u$ generated before and after unlearning, 
including output, 
intermediate parameters, 
embeddings and so on. 
Although the reasons for the disclosure of $D_f$'s information stem from the differences between two versions of the models, 
the information used by attackers is not completely the same. 
Therefore, 
it is crucial for future research to explore how to securely publish both the learned and unlearned models, 
as well as protect the privacy of model components, 
to design effective defense methods.

Additionally,
attribute inference attack and model inversion attack tend to infer the detailed information about $D_f$.
Specifically,
due to the changes of feature representation caused by attribute removal,
attribute-level unlearning may be more vulnerable to attribute inference attack,
while model inversion attack is more likely to affect training data for images (e.g. facial recognition),
or classes with high distinctiveness (e.g. class-level unlearning).
Thus,
further investigation in privacy protection schemes is needed for unlearning in practical applications.

It should be noted that existing research \cite{Wei2023} has confirmed the vulnerability of various types of unlearning techniques to backdoor attacks. 
This vulnerability is caused by data deletion operations, which can trigger backdoors through deleting the preset mislabeled data to cause the unlearned models to generate misclassification,
thereby degrading the model's integrity.
Meanwhile,
to achieve a certain degree of indistinguishability,
approximate unlearning triggers the retraining of model when the approximation errors accumulate within a range.
This provides a vulnerability for poisoning attackers attempting to break the unlearning efficiency.
When considering the integrity and availability of unlearned models,
adversarial attacks may raise security concerns associated with unlearning,
making them a subject worth studying.

In conclusion, 
machine unlearning is not a perfect privacy protection method. 
On the one hand, 
the differences between the learned and unlearned models generated by unlearning may contain sensitive information about the forgotten data. 
On the other hand, 
Backdoor attack and poisoning attack can disrupt the utility of resulting models through unlearning.
Therefore, 
for these newly emerged privacy issues, 
researchers should design unlearning solutions that balance the security and model performance.
A summary and comparison of privacy threats of machine unlearning are provide in Table \ref{comparisonattack}.

\section{Machine Unlearning Applications with Privacy Threats}\label{app}

\subsection{Unlearning in Large Language Models}

Large language models (LLMs) are a typical type of language model,
which utilizes deep learning technologies and large datasets to comprehend,
summarize,
generate and predict content \cite{Min2024},
as well as to answer questions and complete other language-related tasks,
such as ChatGPT \cite{Brown2020}.
Recently,
LLMs have demonstrated superior performance in many NLP tasks,
like language translation and text summarization \cite{Borisov2023}.
However,
these models may also face challenges,
one being copyright infringement \cite{Carlini2021},
where the training set contains data protected by copyright and model generates copyrighted outputs.
This situation may require the removal of such data to prevent privacy risks.
Another one is potential bias issues,
meaning that the training data may have biases that would be reflected through the model generation.
Several research attempted to solve the above problems through machine unlearning techniques.

Jang et al. \cite{Jang2023} proposed a knowledge unlearning method as an efficient solution to privacy issues in LLMs.
It is a post-processing method,
which reduces the model's predictive capability on the forgotten data by altering the direction of gradient updates,
thereby mitigating the privacy risks.
They changed the original training objective from minimizing the negative log-likelihood of target data
to maximizing the negative log-likelihood value.
They updated model parameters using gradient ascent to decrease the prediction probability on the unlearned data.
In addition,
to verify whether the target data has been forgotten,
they also proposed the threshold to quantify privacy risks,
including extraction likelihood and memorization accuracy.
After each iteration,
if the prediction probability of data falls below a certain threshold,
it indicates successful unlearning.
Knowledge unlearning in scenarios that are vulnerable to extraction attacks can provide stronger privacy guarantees, 
as well as increased efficiency and robustness.

Eldan et al. \cite{Eldan2023} proposed a method for selective unlearning specific paragraphs in LLMs.
This method trains the baseline model using reinforcement learning to identify marks of target content.
Then,
it replaces the expressions in the target data and generates labels of each mark based on model prediction.
Finally,
fine-tuning these labels to eliminate the original information of marks in the model.
This method can effectively implement data removal in LLMs,
and has the great potential to solve copyright issues.

Pawelczyk et al. \cite{Martin2023} formalized a new unlearning paradigm for LLMs,
and proposed a black-box removal mechanism,
In-Context Unlearning (ICUL),
aiming to unlearn data points by providing specific contextual information during the inference stage,
without having to update model parameters.
This method flips label on the forgotten data to eliminate its influence on the model,
and then adds correctly labelled training samples to input to mitigate the impact of the label flipping operation.
Finally,
the constructed context with the query input of forgotten data is fed to LLMs.
ICUL provides competitive model performance on real-world datasets and exhibits strong robustness against membership inference attacks.
However,
such black-box unlearning method introduces higher computational burdens.

Large language models exist with harmful social biases,
which may lead to unfairness in natural language processing procedures.
Yu et al. \cite{Yu2023} proposed a Partitioned Contrastive Gradient Unlearning (PCGU) to identify the sources of problematic inferences in the model,
and systematically retrain those parts of model to unlearn biased data.
The basic idea of PCGU is to optimize the weights that makes the greatest contribution to a particular bias domain by comparing gradients of sentence pairs.
Specially,
the gradient is computed for a pair of sentences whose difference is in the bias domain,
and the rank of weights is calculated by using a gradient-based importance algorithm.
Then,
with the gradients and ordered weights as inputs,
PCGU computes a first-order approximation of bias gradient to optimize LLM.
This method is highly effective in mitigating social bias in LLMs while also lowering costs.


\subsection{Unlearning in Federated Learning}

Federated Learning(FL), first proposed by Mcmahan et al. \cite{Mcmahan2017}, 
is a decentralized computing technology capable of training machine learning models through independent local clients with a centralized server. 
In this architecture, 
each client independently conserves its data which is inaccessible to other clients or the central server. 
The central server aggregates only the clients' model parameter updates to train the global model \cite{Konevcny2016federated,Konevcny2016}. 
This design necessitates extensive iterative training to store parameter updates from each local client.
Owing to such training protocols, 
it is challenging to extend centralized machine unlearning frameworks, like SISA, to a federated learning setting \cite{Liu2020}.

Existing research on federated unlearning primarily emphasizes eliminating the historical contributions of a particular client, 
acknowledging that erasing the historical parameter updates could adversely affect the global model. 
To tackle this problem, several federated unlearning methodologies have been proposed. 
Liu et al. \cite{Liu2020} firstly developed an effective federated unlearning method known as \emph{FedEraser}.
Specifically, 
federated learning procures parameter updates through the local client training, forming the global model. 
The central server retains parameter updates and the corresponding indexes from each client and round, enabling the unlearning model to be reconstructed through the adjustment of preserved updates.
This federated unlearning algorithm effectively erase client data and its influence on the global model while decreasing computational time. 


Gong et al. \cite{Gong2022} proposed a particle-based Bayesian federated unlearning framework known as Forget-Stein Variational Gradient Descent (SVGD). 
This framework performs the SVGD updates at the local client whose data is to be deleted, 
employing non-parametric Bayesian approximate inference \cite{Liu2016} and distributed SVGD \cite{Kassab2022}, 
thus achieving high unlearning efficiency. 
Che et al. \cite{Che2023} proposed a fast federated unlearning method.
They first used PCMU scheme \cite{Zhang2022prompt} to train a local unlearning model on each edge device,
and then modified the local unlearning model based on the nonlinear function theory.
The modified local unlearning models are aggregated to generate the global unlearning model,
which provides the certified guarantees.
Pan et al. \cite{Pan2023} solved the problem of unlearning in Federated Cluster (FC),
which introduces a sparse compressed multiset aggregation (SCMA) scheme to aggregate the initial cluster centers generated by each local client using $k$-means.
Zhang et al. \cite{Zhang2023fed} proposed a differentially private machine unlearning algorithm,
FedRecovery.
This algorithm deletes the weighted sum of gradient residuals from the global model,
to eliminate the influence of a forgotten client.
They also provided a rigorous guarantee by adding calibrated noises.
FedRecovery offers competitive model performance and lower computational cost.
To solve the specific requests in FL,
Wu et al. \cite{Wu2022federated} formulated an unlearning framework for three types of deletion requests in FL, namely class unlearning, 
client unlearning, and sample unlearning. 
Likewise,
Wang et al. \cite{Wang2022} highlighted the problem of class-level unlearning in FL. 
They proposed an unlearning algorithm called scrubbing,
to extract specific class information from the learned model, 
quickening the unlearning process without sacrificing accuracy.

\subsection{Unlearning in Anomaly Detection}

Machine unlearning is not necessarily detrimental to the accuracy and overall performance of machine learning models, 
as the potential presence of abnormal samples in the training dataset could actually lead to incorrect predictions and diminished model utility. 
In order to address this issue, 
anomaly detection is used to identify those rare observations that significantly deviate from the majority of samples \cite{Pang2021}. 
Such a method has exhibited importance in varied domains, 
including computer vision, 
financial surveillance, 
and cybersecurity \cite{Bao2019, Georgescu2021}. 
In this context, 
it is noteworthy to examine the value of performing additional abnormal data detection after deploying the machine learning model. 
However, 
a challenge arises during such lifelong anomaly detection; 
typically it is assumed that historic task data becomes inaccessible when new knowledge is acquired. 
Direct application of Stochastic Gradient Descent (SGD) on the model in these circumstances may give rise to the phenomenon known as catastrophic forgetting \cite{French1999}. 
Responding to this problem, 
several research initiatives have proposed the use of machine unlearning as a solution for both anomaly detection and catastrophic forgetting in order to ensure the consistency of the model's superior utility.

Du et al. \cite{Du2019}, 
for instance, 
devised an unlearning method specifically to address the problem of anomaly detection. 
The core idea of this approach is the elimination of harmful samples, 
including false negatives and positives, from the model, 
followed by updating the model in such a way that its performance is retained post-unlearning. 
This method has been empirically validated to be applicable to the majority of 0-1 deep learning-based anomaly detection algorithms, 
including Gaussian mixture models, 
Long Short-Term Memory, 
and autoencoder, 
successfully transforming various machine learning models into their corresponding lifelong anomaly detection schemes.

Similarly, Parne et al. \cite{Parne2021} examined the benefit of unlearning for anomaly detection under lifelong settings, 
utilizing incremental models such as Naive Bayes and Decision Trees. 
Their approach enables fast and comprehensive removal of contaminated data. 
Liu et al. \cite{Liu2022} proposed the formal concept of machine unlearning in lifelong anomaly detection, 
and a \emph{CLPU} unlearning framework on the basis of \emph{SISA} \cite{Bourtoule2021}.
The \emph{CLPU} framework provides a mechanism for sequentially learning a series of tasks, 
enabling exact data removal by constructing a temporary isolated network. 
This framework categorizes sequential tasks into three specific problem types:
(a) tasks to be permanently learned,
(b) tasks to be temporarily learned and later forgotten,
(c) tasks to be completely forgotten.
In experiments, 
the \emph{CLPU} framework demonstrated efficient reduction in storage cost and enhancement in the model's efficacy.

\subsection{Discussion and Privacy Vulnerability of Unlearning Applications}
\begin{table*}[h] \scriptsize
		\centering
		\caption{Summary and Comparison of Unlearning Applications}
		\label{comparisonapp}
		\begin{threeparttable}
			\resizebox{\textwidth}{!}{
			\begin{tabular}{cc|cccccc}
				\hline
    \toprule
				\multicolumn{2}{c|}{\makecell[c]{Unlearning \\ Applications}} & \makecell[c]{Unlearning \\ Techniques} & \makecell[c]{Request \\ Types} & Accuracy & Effectiveness & Efficiency & \makecell[c]{Privacy \\Vulnerability} \\ 
    \midrule
				
				
				
				

                \multirow{4}{*}{\makecell[c]{Large \\ Language \\Models}} & 
				\makecell[c]{Jang et al. \cite{Jang2023}} & \makecell[c]{Model Reset} & \makecell[c]{Batch \\ and stream} & \ding{55} & \ding{55} & \ding{51} & \ding{51} \\ \cline{2-8}

                & \makecell[c]{Eldan et al. \cite{Eldan2023}} & \makecell[c]{Model Modification} & Class & \ding{55} & \ding{55} & \ding{51} & \makecell[c]{Membership inference attack*} \\ \cline{2-8}    
				
				& \makecell[c]{Pawelczyk et al. \cite{Martin2023}} & \makecell[c]{Data Modification} & Sample & \ding{55} & \ding{55} & \ding{51} & \ding{51} \\ \cline{2-8}
				
				& \makecell[c]{Yu et al. \cite{Yu2023}} & \makecell[c]{Model Modification} & \makecell[c]{Sample \\ and batch} & \ding{55} & \ding{55} & \ding{51} & \makecell[c]{Poisoning attack*}\\ \cline{1-8}
    
				\multirow{6}{*}{\makecell[c]{Federated \\ Learning}} &
				\makecell[c]{Liu et al. \cite{Liu2020}} & \makecell[c]{Model Reset} & Client & \ding{55} & \ding{55} & \ding{51} & \ding{51}\\ \cline{2-8}
				
				& \makecell[c]{Gong et al. \cite{Gong2022}} & \makecell[c]{Model Modification} & Client & \ding{51} & \ding{51} & \ding{51} &  \multirow{4}{*}{\makecell[c]{Model inversion attack*}}\\ \cline{2-7}
				
				& \makecell[c]{Wu et al. \cite{Wu2022federated}} & \makecell[c]{Model Reset} & \makecell[c]{Client, sample \\ and class} & \ding{51} & \ding{51} & \ding{51} \\ \cline{2-7}

                & \makecell[c]{Che et al. \cite{Che2023}} & \makecell[c]{Model reset} & Client & \ding{55} & \ding{55} & \ding{51} \\ \cline{2-7}
				
				& \makecell[c]{Pan et al. \cite{Pan2023}} & \makecell[c]{Data partition} & Client & \ding{51} & \ding{51} & \ding{51} \\ \cline{2-8}

                & \makecell[c]{Zhang et al. \cite{Zhang2023fed}} & \makecell[c]{Model Reset} & Client & \ding{51} & \ding{51} & \ding{51} & \ding{51}\\ \cline{2-8}
    
				& \makecell[c]{Wang et al. \cite{Wang2022}} & \makecell[c]{Model Modification} & Class & \ding{51} & \ding{51} & \ding{51} & \ding{51}\\ \cline{1-8} 

				\multirow{3}{*}{\makecell[c]{Anomaly \\ Detection}} & 
				\makecell[c]{Du et al. \cite{Du2019}} & \makecell[c]{Model Reset} & Sample & \ding{55} & \ding{55} & \ding{51} & \multirow{2}{*}{\makecell[c]{Poisoning attack* }} \\ \cline{2-7}
				
				& \makecell[c]{Parne et al. \cite{Parne2021}} & \makecell[c]{Model Modification \\ Model Reset} & \makecell[c]{Sample \\ and batch} & \ding{55} & \ding{55} & \ding{51} \\ \cline{2-8}
				
				& \makecell[c]{Liu et al. \cite{Liu2022}} & \makecell[c]{Data Partition} & Stream & \ding{51} & \ding{51} & \ding{51} & \makecell[c]{Membership inference attack*}\\                
    \bottomrule
    
			\end{tabular}
		}
		\end{threeparttable}
\end{table*}
Current works have shown that unlearning applications are focused on three fields:
large language models,
federated learning,
and anomaly detection.
These applications are mainly based on model-oriented unlearning techniques,
and most of them belong to approximate unlearning.
This is because practical applications require high efficiency as modifying the model can reduce the computational cost.
Moreover,
although accuracy and effectiveness have been improved,
further optimization is still needed,
as well as the applicability of exact unlearning in industry should be investigated in the future.


In terms of potential privacy vulnerability,
although some research has focused on the privacy issues in federated unlearning and LLMs unlearning,
including methods providing robustness guarantees against membership inference attacks \cite{Liu2022, Wang2022, Martin2023}, 
model extraction attacks \cite{Jang2023}, 
and poisoning attacks \cite{Wu2022, Halimi2022},
the privacy risks of unlearning applications still need to be investigated.

For LLMs unlearning,
it usually utilizes model modification methods to erase target data and biased data.
These methods fine-tune model parameters based on computed labels or weights, 
making it vulnerable to model inversion attacks using differences of model outputs,
which may expose the unlearned data or classes.
Therefore, 
unlearning applications should also consider the privacy issues associated with unlearned data and models.

For federated unlearning,
the gradient sharing between local clients and global model,
which stores intermediate parameters,
makes it vulnerable to model inversion attacks. 
In the scenario of anomaly detection,
poisoning attacks could be more easily conducted,
as it aims to identify and remove harmful samples using unlearning methods.
When the unlearned data trigger pre-injected backdoors,
it can potentially result in model retraining. Table \ref{comparisonapp} presents a summary of unlearning applications,
as well as the potential privacy vulnerability.

\section{Discussion and Future Directions}\label{future}

This section will summarize the current works on machine unlearning and analyze the potential research trends from the perspectives of design, vulnerability, evaluation, and application.
Meanwhile,
we list several future research directions that need to be addressed in machine unlearning.

\subsection{The Design of Machine Unlearning}

The design of unlearning methods is currently a major research focus.
Machine unlearning not only supports the removal of Euclidean space data,
such as images and text,
but also involves the removal of non-Euclidean space data,
e.g. the graph unlearning scheme proposed by Chen et al. \cite{Chen2022}.
For deletion requests, 
existing methods cover sample-level unlearning,
class-level unlearning and attribute-level unlearning \cite{Tarun2021, Warnecke2021}.
However,
they are mostly limited to specific scenarios,
to solve certain special deletion requests or based on a particular model.
Considering the trade-off between efficiency and accuracy in the unlearning process,
most techniques improve deletion efficiency at the expense of model performance.
Only a few models can achieve or surpass the prediction accuracy of the retrained model.
Additionally,
some methods,
like those described in \cite{Golatkar2021, Moon2023},
offer verification mechanisms and robustness testing.
These approaches can effectively defend against membership inference attack,
though this issue is often overlooked in unlearning.

Based on the evaluation criteria we presented earlier,
we found that very few existing unlearning approaches can satisfy all of them.
Therefore,
how to design an unlearning framework that meets all these criteria needs to be explored,
which is one of the key research directions in the future.
Firstly,
unlearning methods require high compatibility,
which should not only be applicable to different deletion requests,
but also easily extendable to different models,
especially models with complex structures,
like DNNs and GANs.
Secondly,
unlearning should strike a balance between performance and efficiency.
Finally,
a good unlearning method should exhibit robustness and be able to defend against attacks.
The unlearning methods combined with other techniques also need to be explored, 
such as using model inversion to construct an approximate training set for few-shot unlearning \cite{Yoon2022}.

\subsection{The Vulnerability of Machine Unlearning}

Machine unlearning should ensure the privacy and security of all training data and models, 
especially the forgotten data.
Existing research has shown that the unlearning operation not only fails to protect the privacy of unlearned data,
but also actually increases information leakage risks \cite{Chen2021when, Wei2023, Hu2023}.
Most unlearning methods have vulnerabilities to privacy attacks.
Backdoor attacks are the most common method against unlearning \cite{Wei2023}. 
Regardless of the type of unlearning techniques, 
when an attacker initiates a malicious unlearning request, 
the hidden backdoor is activated to mislead the resulting models,
which compromises the prediction capability of unlearned models.
Approximate unlearning based on model reset is vulnerable to poisoning attacks due to approximation errors \cite{Marchant2022}.
In addition,
the difference between the outputs of the unlearned model and learned model is the key for adversaries to implement information-stealing attacks to extract the privacy information of unlearned data. 
Specifically, 
data partition techniques like \emph{SISA} \cite{Bourtoule2021} are vulnerable to membership inference attacks due to the differences caused by retraining submodels,
while data modification techniques \cite{Cao2015,Tarun2021} such as adding noise or transforming data are vulnerable to attribute inference attacks that expose attribute information through statistical queries.
Recommendation unlearning \cite{Chen2022recommendation} based on graph may also face the same privacy issues.
In federated unlearning, the input of the model may be inferred from the gradient sharing through model inversion attacks.

Therefore, 
the vulnerability and privacy risks of machine unlearning are the key direction of future research.
The availability of some attacks on unlearning,
like adversarial attacks and model extraction attacks,
have not been explored,
especially for unlearning in generative models and large language models. 
Since generative models can generate realistic image data, 
it is worth focusing on investigating whether a generator's behaviors can be copied or the unlearned data can be reconstructed through attacks after unlearning.
On the other hand,
the design of unlearning should also provide robustness guarantees,
especially for backdoor attacks.
How to apply the possible defense strategies mentioned in Section \ref{possibledefense} in machine unlearning frameworks,
e.g. differential privacy and confidence scores masking,
as well as the design of defense methods against existing attacks,
are the important problems that need to be addressed.
Additionally,
unlearning can also be considered as a new solution to defense against adversarial attacks.

\subsection{The Evaluation of Machine Unlearning}

Evaluation of machine unlearning is indispensible.
General metrics,
like accuracy, effectiveness and deletion efficiency (run time)
are easy to implement,
and provide an intuitive understanding for users.
These common evaluation schemes in existing unlearning research \cite{Becker2022}.
However, 
these metrics cannot give strong verification capability for unlearning.
Although theoretical evaluation provides interpretable analysis of unlearning,
it requires numerous assumptions and is challenging to apply to complex model.
For the evaluation of unlearning in terms of verifiability and robustness,
attack-oriented methods are commonly used,
like membership inference attacks and backdoor attacks \cite{Graves2021, Sommer2020}.
But these methods lack applicability and are difficult to implements.

Therefore,
a unified evaluation criterion needs to be proposed to ensure the actual performance of each unlearning method.
This evaluation scheme should not only provide comprehensive and strong metrics,
but also offer users a simple,
feasible and understandable verification mechanism to determine whether their requests has been completed.

In addition,
it is worth noting that theoretical evaluation provides new solution to the investigation of the interpretability of unlearning.
Interpretability aims to explain the functions of various features during the process of model training,
but in machine unlearning,
it should focus on describing that how a model is impacted by the elimination of particular data sets. 
Designing a theoretical evaluation method based on the influence theory can help verify the effectiveness of the unlearning process,
and test the efficacy of data deletion. 

\subsection{The Applications of Machine Unlearning}

The unlearning techniques offer new solutions to privacy protection issues for individual data in several applications scenarios.
For example,
removing the links between users and items in recommender systems \cite{Chen2022recommendation},
eliminating the influence of data from a certain local client on the global model in federated learning \cite{Liu2020},
and erasing copyrighted training data from large language models \cite{Jang2023}.
At the same time,
unlearning is more than erasing data.
In some practical applications,
efficient unlearning would be beneficial.
For anomaly detection,
unlearning can be used to optimize the training set polluted by poisoning attack or accidental mistakes,
to remove the abnormal samples and improve the overall performance of models \cite{Parne2021}.
For potential social biases in LLMs,
unlearning can effectively alleviate the biases in training data through reducing the impcat of relevant weights on the model,
to solve fairness issues in NLP tasks \cite{Yu2023}.

Therefore,
the enormous potential of machine unlearning in practical applications makes it the key issue that requires in-depth investigation.
On one hand,
model repair can be implemented through machine unlearning.
Abnormal samples in the training set may harm the model performance \cite{Liu2022backdoor, Wang2019neural},
while removing such data can repair the model,
which can be seen the process of unlearning \cite{Chen2023turning}.
On the other hand,
unlearning can help to eliminate biased data,
to address model fairness issues.
Lastly,
the applications of unlearning in large models (e.g. LLMs) and generative models should be an important topic in relevant fields.

\section{Conclusions}\label{conclusions}

Machine learning has become an indispensable part of innovation and development in a wide range of fields.
Due to privacy,
the right to be forgotten and other legal requirements,
users may request the removal of individual data and its influence from machine learning models.
As a new solution to data protection in machine learning,
machine unlearning has been conducted in many studies.
However,
current works have not fully provided the explanation of potential privacy risks in machine unlearning schemes.
In this survey,
we provided a comprehensive overview of machine unlearning techniques with a particular focus on new emerged privacy risks.
Firstly,
we clarified the concept and fundamental elements of machine unlearning,
proposed a taxonomy of unlearning schemes (data-oriented and model oriented techniques),
and analyzed and compared the advantages and limitations of current unlearning methods.
Then,
we focused on the importance of privacy risks caused by unlearning,
and reviewed potential attack methods on unlearning.
We also analyzed the vulnerability of existing unlearning approaches.
In addition,
we reviewed the applications of unlearning,
like federated learning and large language models,
as well as their privacy threats.
Finally,
we discussed the research frontiers and development trends,
and provided the feasible directions that need to be further investigated in the future.


\bibliographystyle{unsrt}
\bibliography{manuscript-acmsmall}

\appendix


\end{document}